\documentclass[prb,twocolumn,showpacs,floatfix]{revtex4-1}
\usepackage{graphicx}% Include figure files
\usepackage{dcolumn}% Align table columns on decimal point
\usepackage{bm}% bold math
\usepackage{color}

\begin{document}
\title{Hyperfine interaction and electronic spin fluctuation study on Sr$_{2-x}$La$_x$FeCoO$_6$ (x = 0, 1, 2) by high-resolution back-scattering neutron spectroscopy}
\author{T. Chatterji$^1$,  B. Frick$^1$, M. Zamponi$^2$,  M. Appel$^{1}$, H. S. Nair$^3$, R. Pradheesh$^{4}$, G. R. Hariprya$^{4}$, V. Sankaranarayanan$^4$ and K. Sethupathi$^4$ }
\address{$^1$Institut Laue-Langevin, 6 rue Joules Horowitz, BP 156, 38042 Grenoble Cedex 9, France\\
$^2$J\"ulich Centre for Neutron Science at Heinz Maier-Leibnitz Zentrum, Forschungszentrum J\"ulich GmbH, D-85748 Garching, Germany\\
$^3$University of Texas El Paso, 500 W. University Avenue, El Paso, TX 79968, USA\\
$^4$Department of Physics, Low Temperature Physics Lab, Indian Institute of Technology Madras, Chennai
600036, India }
\date{\today}

\begin{abstract}
The study of hyperfine interaction by high-resolution inelastic neutron scattering is not very well known compared to the other competing techniques viz. NMR, M\"ossbauer, PACS etc. Also the study is limited mostly to magnetically ordered systems. Here we report such study on  Sr$_{2-x}$La$_x$FeCoO$_6$ (x = 0, 1, 2) of which first (Sr$_2$FeCoO$_6$ with x = 0) has a canonical spin spin glas, the second (SrLaFeCoO$_6$ with x = 1) has a so-called magnetic glass and the third (La$_2$FeCoO$_6$ with x = 2) has a magnetically ordered ground state. Our present study revealed clear inelastic signal for SrLaFeCoO$_6$, possibly also inelastic signal for  Sr$_2$FeCoO$_6$ below the spin freezing temperatures $T_{sf}$ but no inelastic signal at all for for the magnetically ordered La$_2$FeCoO$_6$ in the neutron scattering spectra. The broadened inelastic signals observed suggest hyperfine field distribution in the two disordered magnetic glassy systems and no signal for the third compound suggests no or very small hyperfine field at the Co nucleus due to Co electronic moment.  For the two magnetic glassy system apart from the hyperfine signal due only  to Co, we also observed electronic spin fluctuations probably from both Fe and Co electronic moments.
\end{abstract}

 \maketitle 
\section{Introduction}
The study of hyperfine interaction by high resolution inelastic neutron scattering \cite{heidemann_70} is well established by now. However, this technique is much less familiar with the science community compared to the M\"ossbauer, NMR, $\mu$SR and other techniques. The inelastic neutron spin flip scattering probes the hyperfine field at the nucleus and is limited by the cross section and nuclear spin to a certain number of magnetic atoms of which a handful have been studied \cite{heidemann_CoP,heidemann72,chatterji00,chatterji12,chatterji10,chatterji13,chatterji13a} up to now, like V, Co, Nd and Ho. Similar selection limits do also exist for the competing M\"ossbauer and NMR techniques.

So far the study of hyperfine interaction by high energy resolution neutron scattering was mainly applied to magnetically long-range ordered materials. To our knowledge structural disorder was addressed rarely except in an early study of ferromagnetic amorphous CoP$_x$ alloys, were the influence of occupational disorder in the nearest neighbour shell of Co on the hyperfine field split spectra was studied \cite{heidemann_CoP}. Whereas hyperfine splitting (hfs) show up in high resolution backscattering as a resolution limited triplet peak structure centred symmetrically around zero energy transfer, it was shown in this study that a phosphorous concentration dependent distribution of hyperfine fields leads to a broadening of the inelastic excitations observable with neutron backscattering.  

We investigate here two other categories of magnetically disordered materials, a canonical spin glass and a so-called magnetic glass by high resolution inelastic neutron backscattering. The existence of hfs in short-range ordered magnetic systems is not clear. Naively one could even think that one could get no inelastic signal perhaps from such materials due to an absent or zero averaged local field or that one could expect a quasielastic-like signal, which may arise due to extreme field distributions induced at the nuclei from the magnetic ions having different environments (a similar observation of quasielastic-like scattering instead of sharp inelastic lines was made on disordered systems with rotational tunnel splitting \cite{tunneling}). This may be the case for some disordered magnets. Also since this technique essentially probes the magnetic field at the nucleus due to the ordered electronic moment, one expects for ordered systems to see an inelastic hfs signal as soon as the atomic spins are frozen. At higher temperatures, where the electronic spins are mobile and the field at the nucleus averages to zero, we expect for disordered magnets like for the magnetically ordered counterpart no hfs signal. But if the electronic spins of such disordered magnets  freeze below a certain temperature, named spin freezing temperature T$_{sf}$, we may expect to see hfs originating from a residual magnetic moment at the nuclear site. However the inelastic signal may possess finite broadening due to field distributions\cite{heidemann_CoP}. 

Apart from hfs inelastic neutron scattering is known to detect electronic spin fluctuations as quasielastic broadening of the elastic line if their electronic spin fluctuation or relaxation time $\tau_r$ is shorter than the corresponding spectrometer resolution. For the neutron backscattering spectrometers with better than 1$\mu$eV full width energy resolution discussed here this might be detected for $\tau_r$ being shorter than about 4 ns. For multi-atomic samples as studied here the quasielastic scattering can potentially arise from spin fluctuations in the electronic environment of any atom.

Double perovskite materials A$_2$BB$^{\prime}$O$_6$ (A = Alkaline earth or rare earth ions; B,B$^{\prime}$ = transition metal ions) have drawn intense interest in the condensed matter and materials science community. There are several interesting aspects about these materials viz. antisite disorder, competing magnetic interactions and transition metal spin-state transitions. Antisite defects in double perovskites are related with the magnetoresistance observed in these materials \cite{garcia01}. They also contribute to the competing exchange interactions, that lead to frustration and spin glass state \cite {carlo11}. Existence of different spin states of B and B$^{\prime}$ influences Jahn-Teller effect. The  A$_2$BB$^{\prime}$O$_6$ materials are also reported to show multiferroicity \cite{du10} and magnetocapacitance \cite{rogado05} making them truly multifunctional materials. 

We have chosen here the class comprising of the series Sr$_2$FeCoO$_6$, SrLaFeCoO$_6$ and La$_2$FeCoO$_6$. Among these Sr$_2$FeCoO$_6$ has been shown to be a canonical \emph{spin-glass} system \cite{pradheesh12epj,pradheesh12} whereas SrLaFeCoO$_6$ can be termed as the so-called \emph{magnetic glass} \cite{pradheesh17}. The last member La$_2$FeCoO$_6$ becomes \emph{magnetically ordered} \cite{pradheesh18} below about 225 K .
The main aim of our present study concerns the search by neutron backscattering for possible hyperfine splitting (hfs) at the Co-site which gives important local information also concerning the electronic magnetism of different selected systems. Further we investigate the electronic spin fluctuation found in some of these systems in a certain temperature range. The measured spectra and elastic intensity contain also information on how the magnetic intensities develop with temperature, which together with the Q-dependence reveals interesting new information about the electronic magnetism in these systems.

\section{Experimental}  \label{sec:Experimental}
Powder samples of Sr$_2$FeCoO$_6$, SrLaFeCoO$_6$ and La$_2$FeCoO$_6$ were prepared by a sol-gel method \cite{pradheesh12} and about 3 g of material were placed in Al sample holders which were fixed either to the cold tip of the top loading closed-cycle cryostat or in a cryofurnace. %\textcolor{blue}{\emph{add temperature history? seems to be important for some samples}.}

High resolution inelastic neutron scattering experiments were carried out on two different backscattering spectrometers, at SPHERES \cite{wuttke} operated by  J\"ulich Center for Neutron Science at the MLZ in Garching, Germany and at the backscattering spectrometer IN16B \cite{in16b,in16bexperimentDOI}  at the Institut Laue-Langevin, Grenoble. On both instruments the wavelength of the incident neutrons is $\lambda = 6.271$ {\AA} with an energy resolution of FWHM $\approx$ 0.7 $\mu eV$ in their standard configurations with strained Si 111 backscattering crystals and the maximum energy transfer lies near 30 $\mu eV$. For most measurements on both instruments the energy range was deliberately restricted to the range where the hyperfine splitting is expected thus optimising the count rate. Only on IN16B some samples were measured over the full energy transfer range to check for quasielastic scattering. 

On IN16B different instrument configurations were used for the three experiments carried out. In addition to the standard configuration a part of the  samples was measured with either the higher energy resolution but lower flux ('polished Si 111' configuration \cite{in16b} ) or a 'high signal-to-noise ratio' mode (HSNR) \cite{appel}. The SrLaFeCoO$_6$ sample was measured in the standard high flux configuration (FWHM $\approx$ 0.75  $\mu eV$) and in addition at a few temperatures in the HSNR mode. Sr$_2$FeCoO$_6$ and at two temperatures also La$_2$FeCoO$_6$ and SrLaFeCoO$_6$ were measured with high resolution 'polished Si 111 setup' \cite{in16b} with 'polished' analysers in the low Q-range below 1.06 {\AA}$^{-1}$ for which an energy resolution of FWHM $\approx$ 0.31  $\mu eV$ was obtained. The lower flux intrinsic to the higher energy resolution and the reduced analyser surface lead to a considerably lower count rate compared to the standard IN16B configuration. For the same experimental setup  ÒstrainedÓ Si 111 crystal analysers were placed at the larger scattering angles (Q $\gtrsim$ 1.06 {\AA}$^{-1}$) for which a resolution of FWHM $\approx$ 0.6  $\mu eV$ resulted.
On IN16B both full spectra and 'elastic fixed window scans' (efws) were carried out. In efws only neutrons scattered without energy change are counted as function of temperature. IN16B spectra were corrected with LAMP \cite{lamp} for background and self-screening and were normalised to Vanadium. 

On SPHERES for all samples only quasielastic spectra were measured in the same standard setup within an energy transfer of $\pm$5  $\mu eV$ and an energy resolution of FWHM $\approx$ 0.6  $\mu eV$.

\section{Results}
Out of the three samples measured on two back-scattering spectrometers only the \emph{magnetic glass} sample, SrLaFeCoO$_6$ \cite{pradheesh17}, showed at low temperatures the inelastic signal expected for hfs of the Co nuclear ground state and this is why we will present first the results for this sample even though its electronic magnetic state seems to be more complex \cite{pradheesh17}. We observe for SrLaFeCoO$_6$ clear inelastic peaks arising from the hfs which move with increasing temperature towards the central elastic line and finally merge with it near the spin-freezing temperature T$_{sf}$ = 75 K. 

The second sample, the canonical \emph{spin glass} Sr$_2$FeCoO$_6$, showed as well magnetic excitations at low temperatures, however of less clear nature. This is why we have measured this sample extensively in different instrument configurations to decide if the observed scattering for this sample is quasi-elastic or inelastic. 

Curiously the last sample, La$_2$FeCoO$_6$, which seems to be \emph{magnetically ordered} below about T = 225 K, as evidenced from neutron diffraction experiments \cite{giveReference}, did not show any measurable inelastic nor quasielastic signal in the whole temperature range up to 300 K. 

\subsection{Magnetic glass SrLaFeCoO$_6$}

\begin{figure}
\resizebox{0.45\textwidth}{!}{\includegraphics{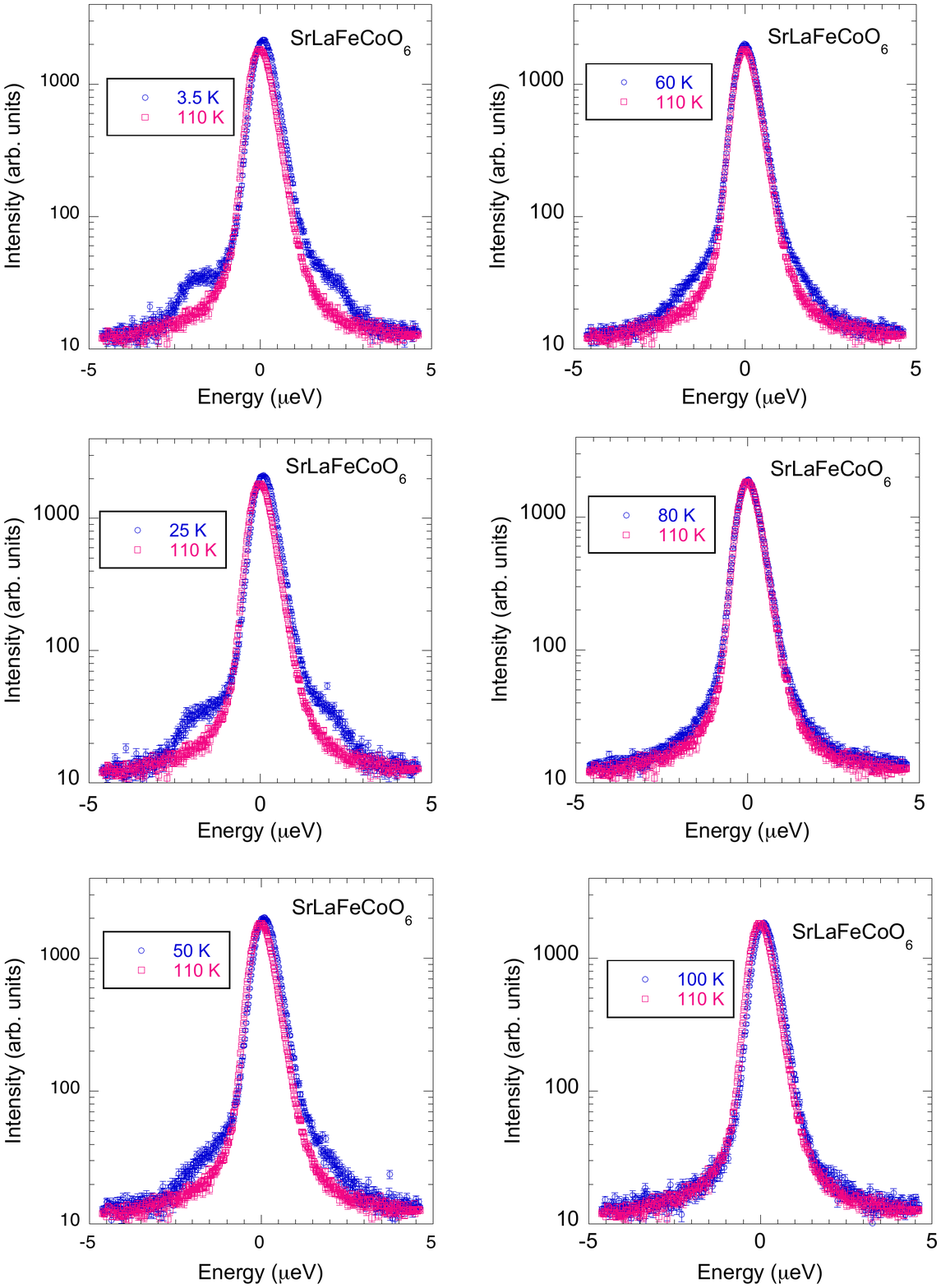}}
\caption {Neutron spectra from SrLaFeCoO$_6$ at several temperatures compared to the spectrum at T = 110 K. There exist extra inelastic scattering at low temperatures below the spin temperature $T_{sf} \approx 75$ K. The inelastic scattering has energy of about 1.8 $\mu$eV which moves towards the central elastic peak as a function temperature and merge to it at about the spin-freezing temperature $T_{sf} \approx 75$ K.
 }
 \label{SLFCO_SPHERES}
\end{figure}

According to neutron powder diffraction SrLaFeCoO$_6$ has a monoclinic P21/n crystal structure, which is retained without phase transition down to 4 K, yet with some weak changes in the temperature dependence of the unit cell volume near T = 250 K and 75 K. No long-range magnetic order is observed in the neutron powder diffraction patterns\cite{pradheesh17}. However, unusual features in magnetization measurements ressemble a kinetic arrest near $T_{sf} \approx 75$ K, which led Pradheesh et al. to classify SrLaFeCoO$_6$ as a so-called \emph{magnetic glass} \cite{pradheesh17} below the spin freezing temperature $T_{sf} $.

From our backscattering spectroscopy on SrLaFeCoO$_6$ we show first the spectra measured on SPHERES at several temperatures in comparison to a spectrum measured at T = 110 K which serves as resolution (see Fig.\ref{SLFCO_SPHERES}).  Below the spin freezing temperature $T_{sf} \approx 75$ K exists extra inelastic scattering, located at an energy of about 1.8 $\mu$eV at T = 3.5 K, which moves with increasing temperature towards the central elastic peak and finally merges with it near $T_{sf}$. The temperature dependence of the inelastic peak positions was determined from different fit procedures explained in the Appendix below, where also some corresponding fit curves are shown (see Appendix Fig. \ref{SLFCO-fit}). Without attributing too much significance to the functional, the temperature dependence of the average hyperfine splitting (shown in Appendix Fig. \ref{SLFCO-fit}) can be parameterised as  power laws with fit parameters which slightly depend on the way the spectra are fitted. Correspondingly magnetic transition temperatures of $T_{sf}$ = 71.1$ \pm 0.8$ K and  $T_{sf}$ = 83 $\pm 1$ K were obtained. 

\begin{figure}
\resizebox{0.5\textwidth}{!}{\includegraphics{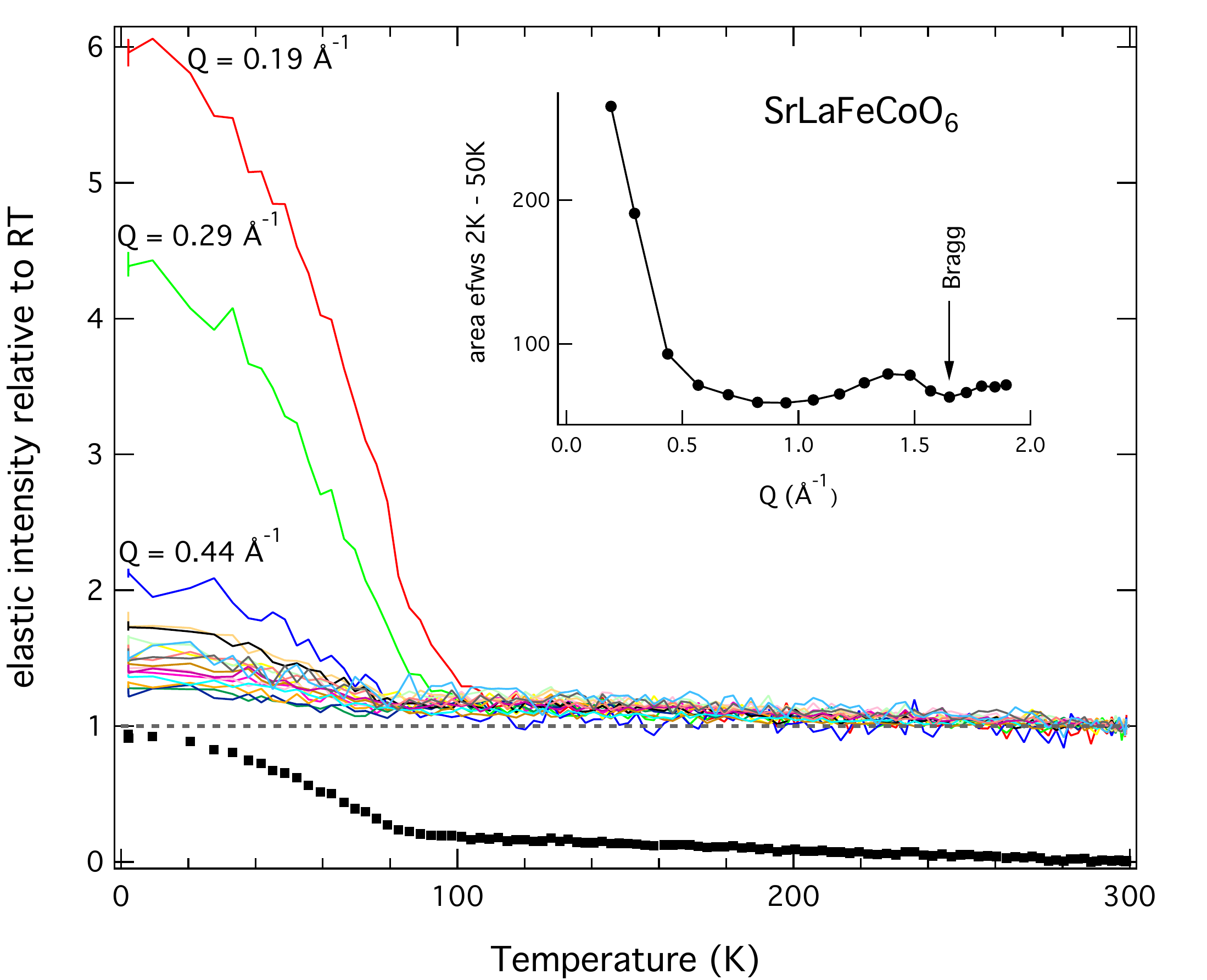}}
\caption {Temperature-dependence of the elastic intensity of SrLaFeCoO$_6$ normalised to its high temperature average value. Main panel, bottom (black squares): sum over all Q-values; for clarity offset by -1. Lines in main panel: elastic intensity for different Q values, with the 3 lowest Q-values labelled. A strong increase of the elastic intensity is observed at temperatures below T = 100 - 80 K. Inset: The Q -dependence of the area under the efws curves between T = 2 and 50 K. Besides the increase at low Q, a small increase of an elastic diffuse contribution can also be observed around the Bragg peak position at $Q = 1.57$ {\AA}$^{-1}$}.
\label{SLFCOefws}
\end{figure}

On the backscattering spectrometer IN16B we have investigated the temperature dependence of the energy resolved elastic intensity by efws (Fig.\ref{SLFCOefws}). With decreasing temperature the Q-averaged elastic intensity shows below $T \approx 80$ K a pronounced increase as shown in the lower part of the main panel of Fig.\ref{SLFCOefws}. Cooling down from 300K the average elastic intensity increases initially weakly until about 80 K where it mounts by about 60\%. It is also shown for individual Q-values as colored lines in the main panel, which reveals that the elastic intensity increase at low temperature is strongest in the low Q region (Fig.\ref{SLFCOefws}).Thus  e.g. at Q = 0.19  {\AA}$^{-1}$ we observe an increase by nearly 600\% below T = 100 K, whereas for higher Q the onset happens rather around 80 K. The Q-dependence of this low temperature elastic intensity increase is summarised in the inset, where the temperature integrated intensity between 2 K an 50 K is presented. Again this shows clearly that the elastic intensity increase is strongest at low Q where the Q-dependence  might resemble a magnetic form factor. In addition we observe at higher Q, around the first structural Bragg peak, at low temperature diffuse elastic scattering. The reported behaviour of the elastic intensity at low temperatures might be interpreted as the onset of magnetic scattering which is static at the ns-time scale below $T_{sf}$. Thus in contrast to neutron diffraction where no magnetic reflections\cite{pradheesh17} were found, we may observe here signs of magnetic short range order and possibly electronic spin freezing between $T \approx 80 - 100$ K. 

\begin{figure}
\resizebox{0.5\textwidth}{!}{\includegraphics{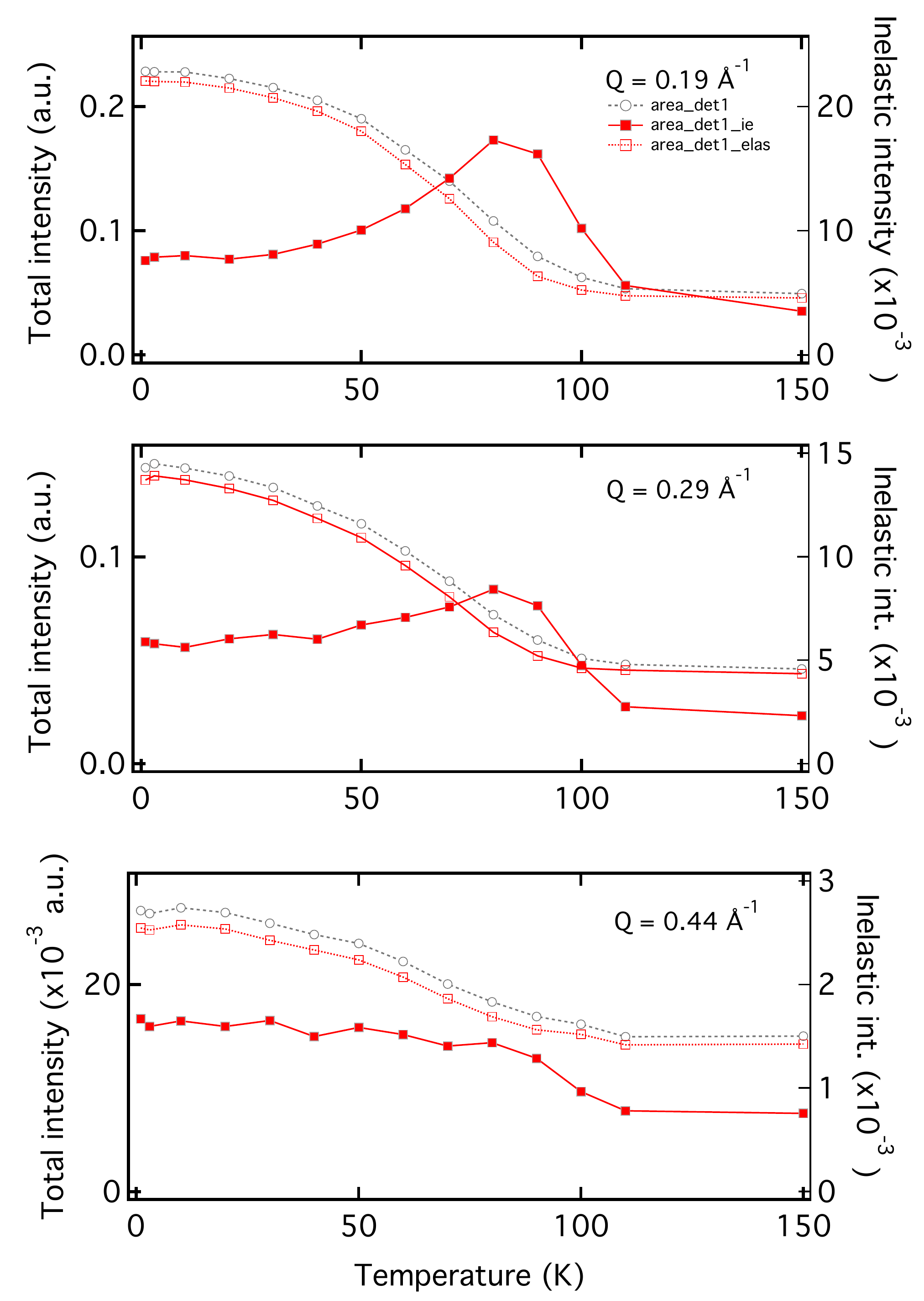}}
\caption {Integrated intensities for SrLaFeCoO$_6$ as deduced from IN16B spectra measured over $\pm$5$\mu$eV for 3 low Q-values. }
 \label{integrate_inelastic_Intensity}
\end{figure}

 Following up the latter hypothesis we investigate now the Q and temperature dependence of energy spectra of SrLaFeCoO$_6$ (an example for spectra measured on IN16B in normal operation mode is given in the Appendix A, Fig.\ref{SLFCOspectra_HFnormal_TdepSums}) for which we detect indeed the onset of quasielastic scattering at low Q when integrating different spectral ranges. The temperature dependence of the intensity in the different energy windows is shown in Fig \ref{integrate_inelastic_Intensity} for the three lowest investigated Q-values. The total intensity (within $\pm$5$\mu$eV) increases with decreasing temperature below 100K very similar to the resolution determined elastic intensity,  which can be interpreted as the onset of magnetic order but is not necessarily a proof for  a change in spin dynamics. In contrast at low Q the inelastic intensity, integrated outside of the resolution wing over the energy range  within 1.5 - 5 $\mu$eV, runs near T = 90 K through a maximum with decreasing temperature and levels then off only around T = 40 K. This behaviour is again most clear at the lowest Q = 0.19 {\AA}$^{-1}$. Such temperature dependence of the inelastic intensity exhibiting within a fixed energy window a maximum is well known and is a signature of a quasielastic signal which has at this temperature its maximum spectral weight in the fixed observation window \cite{frick_ifws}. Very similar but less pronounced behaviour is seen for Q = 0.29 {\AA}$^{-1}$ and at higher Q, shown for Q = 0.44 {\AA}$^{-1}$ where no maximum but a step-like inelastic intensity increase is detected. Thus we conclude that we observe electronic spin fluctuations on the 'ns-time-scale' at low Q or larger distance, which are still visible as quasielastic scattering below $T_{sf} $ and which slow down with decreasing temperature. At higher Q the inelastic intensity does not show a maximum but still a step-like increase, which could mean that at a more local scale the fluctuations are much faster than nanoseconds and slow down at $T_{sf} $. Thus we could in principle as well expect spin dynamics in the high Q-range and freezing into diffuse elastic scattering as has been observed earlier for a frustrated antiferromagnet \cite{mondelli} on IN16.

\begin{figure}
\resizebox{0.5\textwidth}{!}{\includegraphics{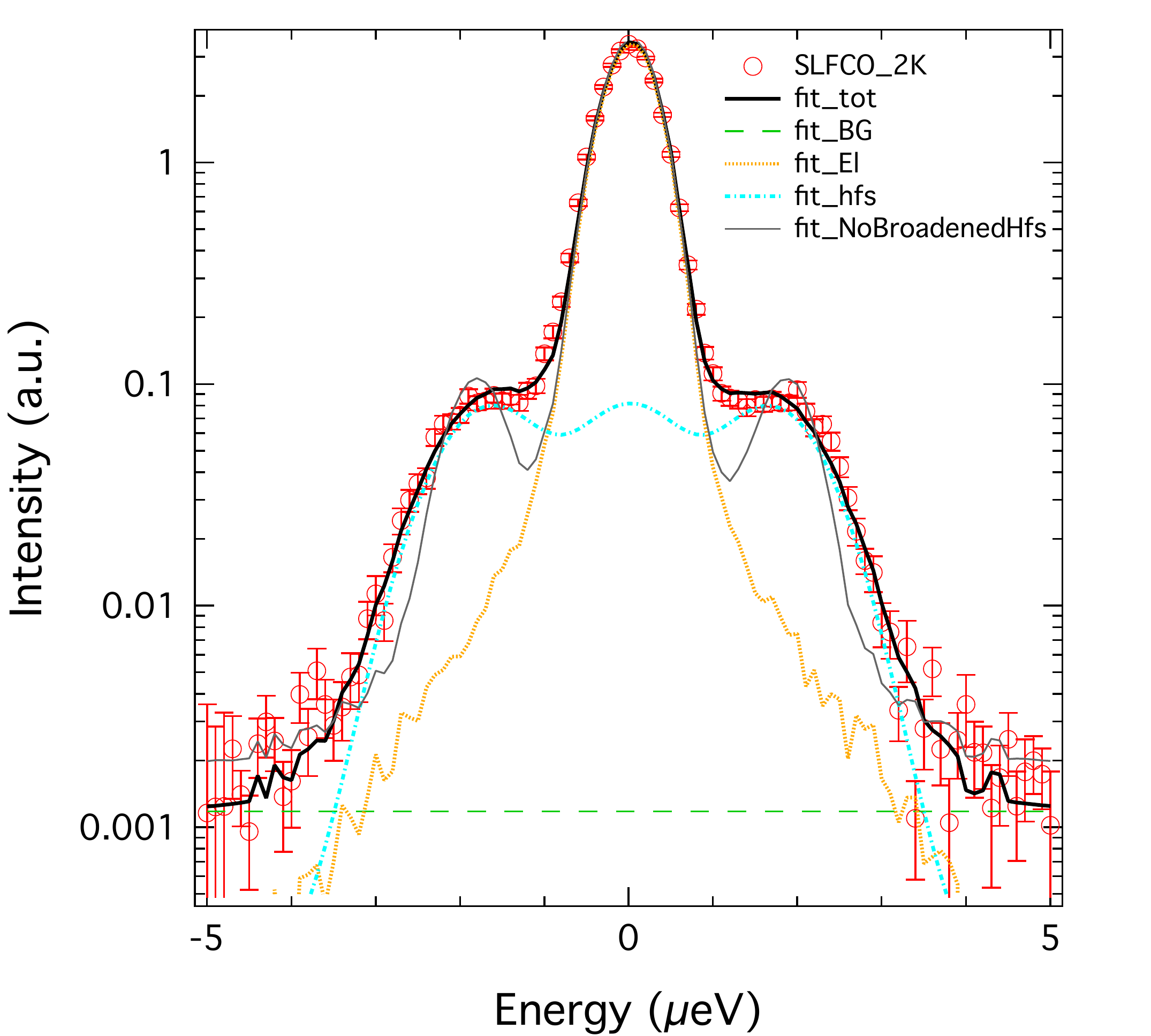}}
\caption {Fit of the SLFCO spectrum at T = 2 K (total fit: thick black line) with a broadened hyperfine spectrum (dash-dotted line), additional elastic component (dotted) and flat background (dashed). Without a broadening of the hyperfine spectrum the total fit shows too sharp features (thin solid line). Experimental data are summed between Q = 0.5 and 1.87  {\AA}$^{-1}$.}
 \label{fit_hfs_SLFCO_2K_ElHfsBg}
\end{figure}

\begin{figure}
\resizebox{0.5\textwidth}{!}{\includegraphics{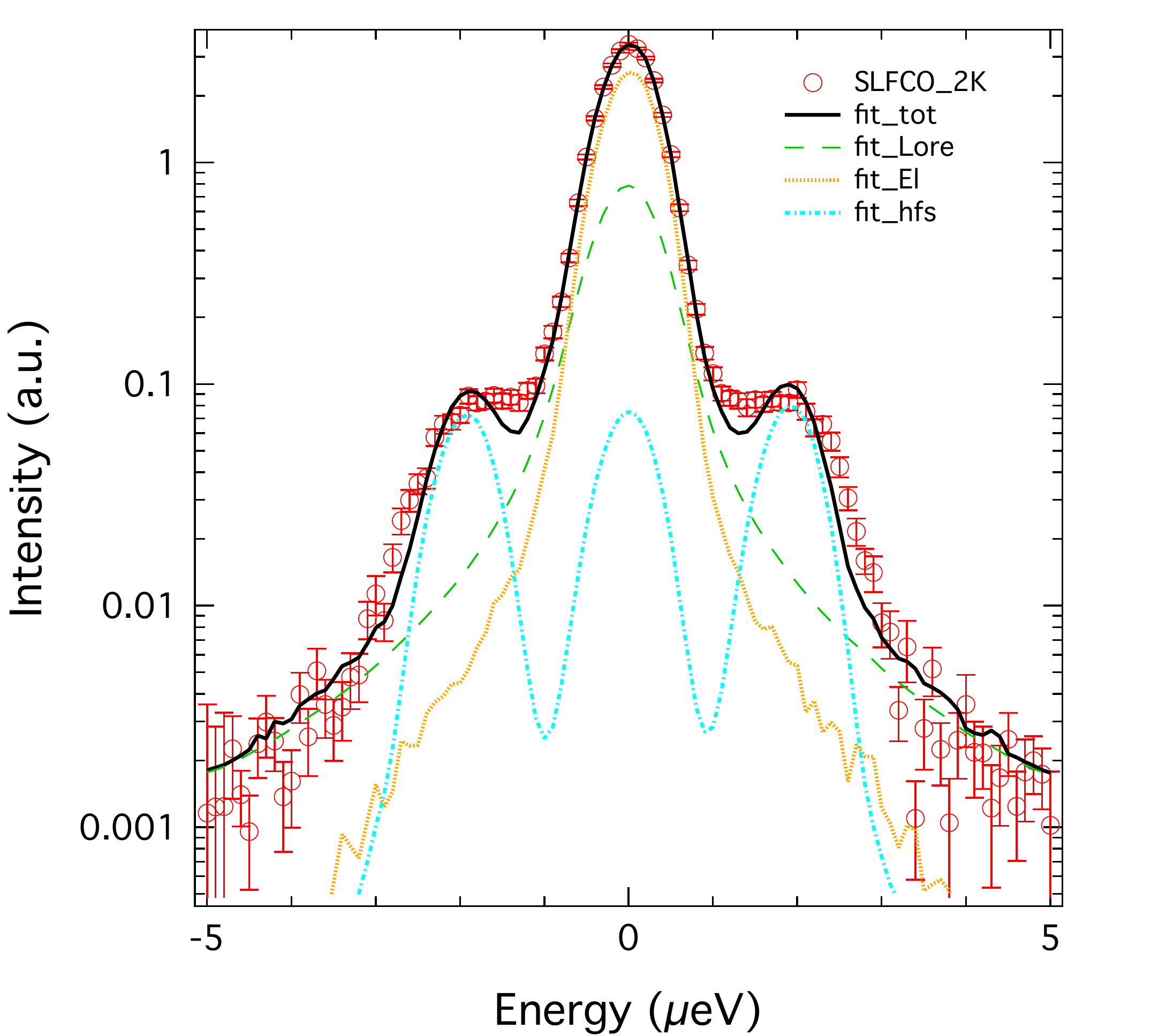}}
\caption {Fit of the SLFCO spectrum at T = 2 K (total fit: thick black line) with an non-broadened hyperfine spectrum (dash-dotted line), elastic component (dotted) and additional quasielastic Lorentzian component (dashed).Experimental data are summed between Q = 0.57 and 1.87  {\AA}$^{-1}$.}
 \label{fit_hfsLorE_SLFCO5b01_2K}
\end{figure}

The observed electronic magnetic ordering and spin freezing is quite certainly at the origin of the residual hyperfine field created at the Cobalt nucleus for which the onset is observed in the same temperature range. When deducing a functional temperature dependence for the hfs at the Cobalt site, as attempted in Fig.\ref{SLFCO-ET}, one should be aware of the possible presence of additional quasielastic scattering, which might influence the data especially near T$_{sf}$. 

Based on these results we have repeated a few measurements on SrLaFeCoO$_6$ with better signal-to-noise ratio using the HSNR mode \cite{appel} of IN16B. The higher sensitivity allows for a more detailed discussion of the model fits of the hyperfine split spectra.  We show in Fig.\ref{fit_hfs_SLFCO_2K_ElHfsBg} and Fig.\ref{fit_hfsLorE_SLFCO5b01_2K} the spectrum and fits for SrLaFeCoO$_6$ at T = 2 K, averaged over higher Q-values only (Q = 0.57 to 1.87  {\AA}$^{-1}$). The spectrum can be fitted well by a single hfs-spectrum (thick solid black line in Fig.\ref{fit_hfs_SLFCO_2K_ElHfsBg};  $\chi^{2}$  = 1.88) with a ground level splitting of 1.67 $\pm$ 0.05 $\mu$eV (consistent with Appendix, Fig.\ref{SLFCO-ET}), an additional elastic contribution as well as an additional weak flat background (2 orders of magnitude lower than the inelastic peak height). Interestingly, for these fits one needs to allow for a broadening of  the hyperfine split lines (Gaussian $\sigma$ $\approx$ 0.715 $\pm$ 0.018 $\mu$eV), because the theoretically expected non-broadened delta function triplet for hyperfine splitting does not describe the spectra well (shown by the thin solid black line in Fig.\ref{fit_hfs_SLFCO_2K_ElHfsBg}). A justification for fitting broadened lines can easily be found in the magnetic shortrange order which is ascribed to site disorder in these perovskites.\cite{pradheesh17} A broadening of hyperfine split lines had also been observed by Heidemann\cite{heidemann_CoP} in early studies of disordered Co alloys. We have also tried to fit with an non-broadened hyperfine spectrum but adding instead a quasielastic Lorentzian component (Fig.\ref{fit_hfsLorE_SLFCO5b01_2K}). Such a fit is somewhat worse ($\chi^{2}$  = 4.5) and results again in somewhat  too sharp hyperfine peaks.

With the HSNR mode we also have attempted to determine how far the observed quasielastic component extends in energy by measuring a few spectra near T$_{sf}$ over a wider energy range ($\pm$30$\mu$eV) at T = 2, 70, 80, 90, 100, 110 and 150K. Integrating these spectra within inelastic energy windows of 5$\mu$eV energy width we find again that the temperature dependent area evidences a maximum near the transition temperature.

Up to here we have demonstrated a very typical nuclear hyperfine field splitting on the Co-site of SrLaFeCoO$_6$, a magnetic system which shows some spinglass like behaviour but otherwise is more complex and freezes below T$_{sf}$ in a so-called 'magnetic glass' phase. In addition we have observed the existence of electronic spin fluctuations which freeze on the ns-time-scale and for high Q near T$_{sf}$ , but only below T$_{sf}$ for low Q, at a more extended spatial scale. Now we will pass to a typical spin glass with less clear hyperfine splitting.

\subsection{Canonical spin-glass sample  Sr$_2$FeCoO$_6$ }  

Neutron powder diffraction and bond valence sums analysis for Sr$_2$FeCoO$_6$ has shown that the B site in this double perovskite is randomly occupied by Fe and Co in mixed valence states of Fe$^{3+}$/Fe$^{4+}$ and Co$^{3+}$/Co$^{4+}$, respectively \cite{pradheesh12}.The resulting competition of ferromagnetic and antiferromagnetic interaction leads to a spin-glass phase with a freezing temperature of $T_{sf} \approx 80$ K \cite{pradheesh12}. Detailed magnetization studies  \cite{pradheesh12} have shown that Sr$_2$FeCoO$_6$ can be considered a canonical spin glass. High resolution neutron diffraction investigation \cite{chatterji} did not  show any appreciable magnetic Bragg scattering down to the base temperature of the cryostat which was about 10 K. 

\begin{figure}
\resizebox{0.5\textwidth}{!}{\includegraphics{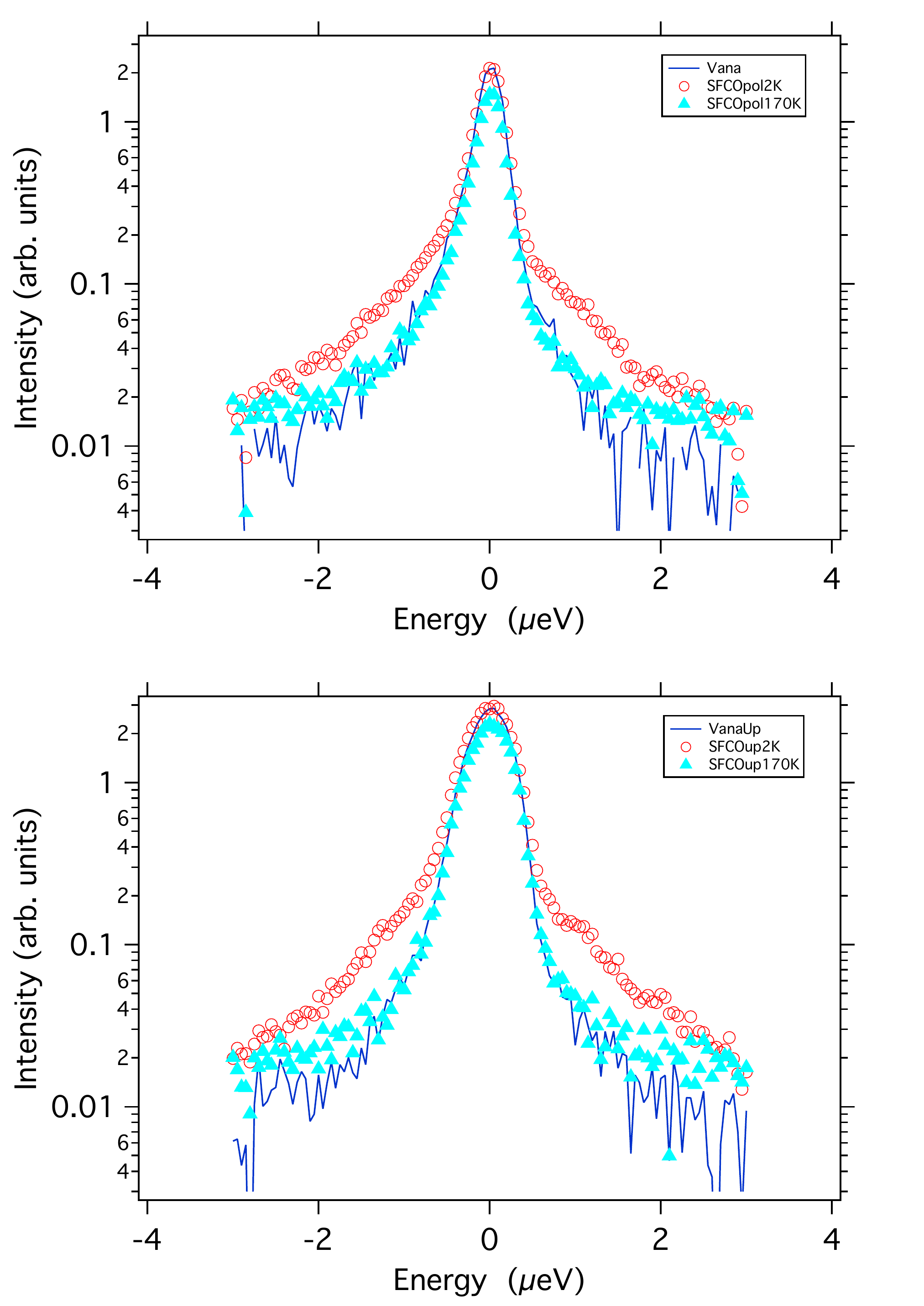}}
\caption {Spectra of Sr$_2$FeCoO$_6$ measured on IN16B with higher energy resolution: a) top spectra summed in the lower Q-range and resolution of FWHM $\approx$ 0.31 \ $\mu$eV and b) bottom: at higher Q-values with the 'unpolished' analysers  (see text) and a resolution of FWHM $\approx$ 0.6 \ $\mu$eV. At T = 2 K one sees clearly the signal from hyperfine splitting, whereas at T = 170 K the line width is identical with the measured Vanadium resolution which is scaled to the elastic peak intensity of the sample at 2 K.
 }
 \label{SFCO2AND170K_comparePolUpol}
\end{figure}

When searching for hyperfine splitting of Cobalt in Sr$_2$FeCoO$_6$ on the back-scattering spectrometer SPHERES, only weak but nevertheless significant indications for extra scattering at T = 3.5 K compared to T = 150 K were found (shown in the Appendix, Fig. \ref{SFCO}), which did not allow to fit the hyperfine splitting. However, these signatures for hfs are clearly confirmed by high resolution measurements on IN16B, when comparing the 2 K spectrum of Sr$_2$FeCoO$_6$ with the Vanadium resolution or with a T = 170 K Sr$_2$FeCoO$_6$ spectrum (see Fig. \ref{SFCO2AND170K_comparePolUpol}). Nevertheless fits need to decide if the observed extra scattering is of inelastic or quasi-elastic nature, which is what we will investigate now in more detail. For a canonical spin glass below the spin-freezing temperature T$_{sf}$ one may expect hyperfine field distribution and therefore a broadening of the inelastic scattering as reported above for SrLaFeCoO$_6$ and for very wide distributions this could eventually even resemble quasi-elastic scattering, similar to what was reported for disordered systems with quantum rotational tunneling \cite{tunneling}.

 The spectra in the top figure of Fig. \ref{SFCO2AND170K_comparePolUpol} are added in the Q-range from 0.44 - 1.06 {\AA}$^{-1}$ and, in order to gain in count rate, were measured in a $\pm$ 3 \ $\mu$eV window only.  The fitted energy resolution for Vanadium was  0.31 $\mu$eV. The spectra in the bottom figure were added in the Q-range from 1.06 - 1.8 {\AA}$^{-1}$ where the measured  energy resolution was $\approx$  0.6 \ $\mu$eV, due to chosen mismatched monochromator and analyser crystals (see section~\ref{sec:Experimental}). At least visually at 2 K the symmetric shoulders on both sides of the elastic line in both figures suggest that there should be inelastic low frequency scattering from hfs. At high temperature, T = 170 K, where the electronic spins fluctuate randomly the  Sr$_2$FeCoO$_6$ spectrum corresponds well to the Vanadium resolution (Fig. \ref{SFCO2AND170K_comparePolUpol}). 

\begin{figure}
\resizebox{0.45\textwidth}{!}{\includegraphics{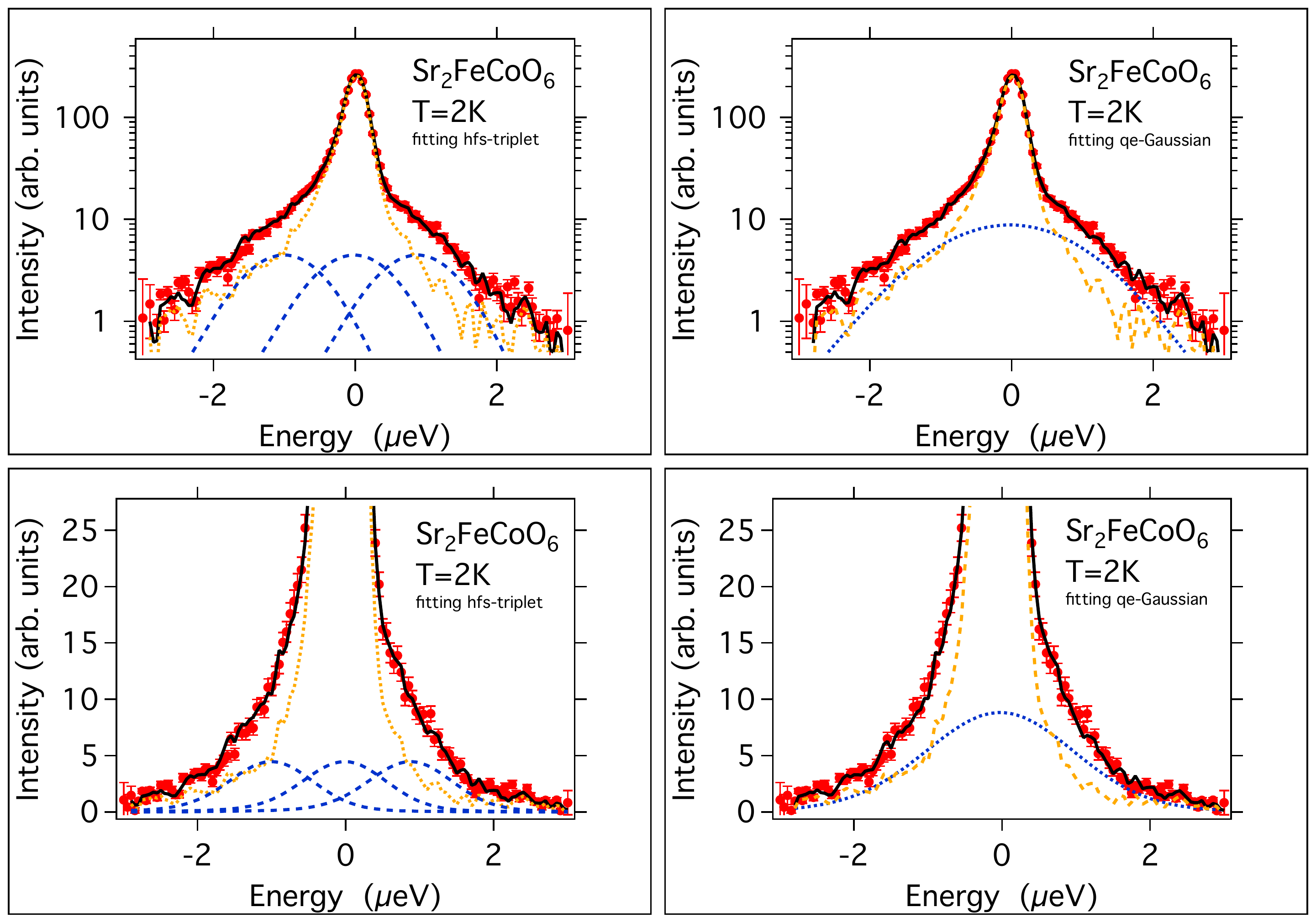}}
\caption {Fits of the spectra of Sr$_2$FeCoO$_6$ at T = 2 K, measured on IN16B with high resolution (FWHM=0.3\ $\mu$eV) ÒpolishedÓ Si111 setup. Spectra were summed between Q = 0.44 and 1.06 {\AA}$^{-1}$. The upper figures are in log scale in intensity and the the lower figures are in linear scale. The black line through the data points (red symbols) is a fit of the total model curve convoluted with the Vanadium resolution function.The dotted yellow line results from a fit of the additional dominating elastic scattering convoluted with the resolution and the blue dashed lines represent the sub-functions used to describe the additional inelastic and quasielastic scattering.
Left column: a typical scattering law for nuclear hyperfine splitting results in a triplet of equidistant (elastic + symmetric inelastic) delta functions of same intensity, convoluted with the instrumental resolution. Here we had to assume in addition a Gaussian broadening of the triplet functions in addition to the convolution with the resolution. Right column: Fit with a single, zero energy entered, Gaussian function convoluted with the resolution function (blue dotted line) and the dominant elastic scattering (yellow dots).}
 \label{LayoutSFCOcompareFits}
\end{figure}

 The higher resolution spectrum corresponding to the upper part of figure \ref{SFCO2AND170K_comparePolUpol}) was fitted in different ways by: A) a typical hyperfine splitting (hfs) consisting of a symmetric triplet of resolution broadened delta functions, B) a quasielastic Gaussian (centred at E=0) and C) a quasielastic Lorentzian (centred at E=0). (see Fig.\ref{LayoutSFCOcompareFits})

\emph{Model A}: A good description by the most plausible hyperfine splitting fit function \cite{heidemann_70} was only possible if additional Gaussian broadening of the hfs peaks due to disorder was assumed. The three Gaussians were constrained to be symmetrically positioned and of equal width and intensity, as shown in Fig.\ref{LayoutSFCOcompareFits} l.h.s. The reasoning behind this fit function is like above for SrLaFeCoO$_6$ that the local induced field at the Cobalt site of the spin glass Sr$_2$FeCoO$_6$ is heterogeneous due to disorder and that the hfs lines are correspondingly broadened. The fits indicate a finite energy splitting which amounts only to about 1 $\mu$eV.

\begin{figure}
\resizebox{0.45\textwidth}{!}{\includegraphics{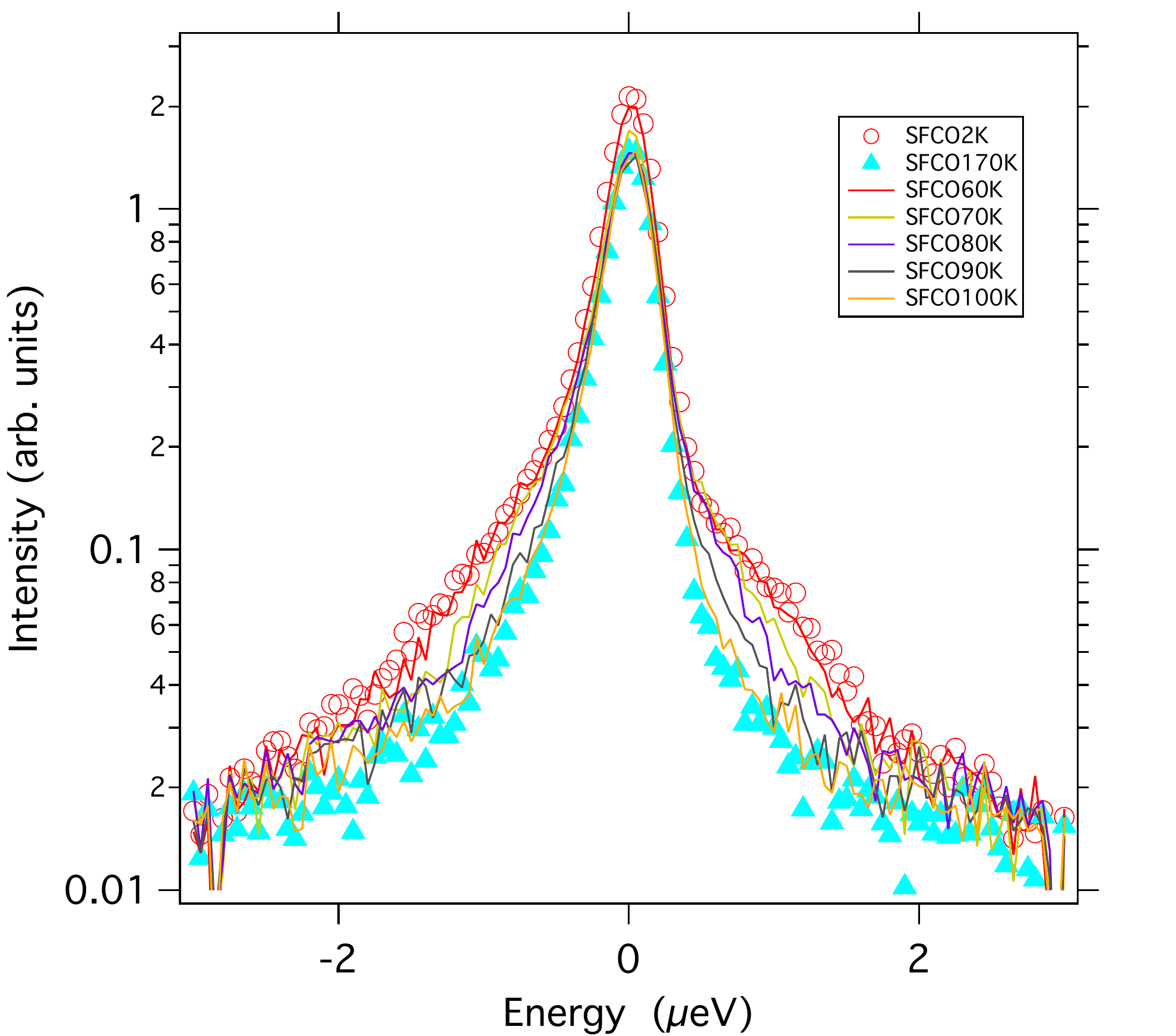}}
\caption {Temperature dependence of the Q-averaged spectra of Sr$_2$FeCoO$_6$ measured on IN16B with higher energy resolution. The temperatures measured between 2K and 60K are not shown for clarity because they are practically unchanged with respect to the 2K spectrum.
 }
 \label{SFCO_Tdep_pol}
\end{figure}

The typical scenario for the temperature dependence of the hfs, i.e. softening of the side peak positions with increasing temperature and finally merging with the elastic line at the magnetic transition temperature was descibed and found above for the magnetic glass SrLaFeCoO$_6$. For Sr$_2$FeCoO$_6$ this is again the observed temperature dependence as can be judged from Fig. \ref{SFCO_Tdep_pol}. For temperatures between 2 K and 60 K the spectra hardly change - this is why intermediate temperatures are not shown for clarity. Then near 70 K the intensity in the wings diminishes as a first sign of softening of the extra scattering compared to low temperatures and finally we observe a merging into the elastic line and the disappearance of the inelastic peak intensity near T$_{sf}$ = 85K or slightly above. For T = 100 K the spectrum appears within statistics similar to the one at 170K. The fit quality with the ÔbroadenedÕ hyperfine splitting model A (shown in Fig. \ref{LayoutSFCOcompareFits} for 2 K) is excellent  for all spectra below T = 70 K with an average $\chi^{2}$ = 1.3 $\pm$ 0.13. The resulting temperature dependence of the fitted side peak positions is shown in the Appendix, Fig \ref{SFCO_hfs_fits_PosInt}.

\emph{Model B}: A fit with a centred quasielastic Gaussian would be a reasonable description of the underlying physics if the local distribution of hyperfine fields and thus the local hfs would average to E = 0. The width of the Gaussian is related to the width of the assumed Gaussian distribution. The fit quality of this model is equally good with an average $\chi^{2}$  factors of $\chi^{2}$  = 1.28 $\pm$ 0.13 below T = 70 K. However, within the investigated energy range the line width of the Gaussian (see Appendix, Fig. \ref{SFCO_qens_fits_WidthInt}) results to be decreasing with temperature. The area of the Gaussian function remains about constant below 70 K and then decreases fast within 20K. In this model a decreasing width would mean that the distribution of the local hyperfine splitting becomes narrower with increasing temperature.  

\emph{Model C}: Fitting a centred quasielastic Lorentzian function could be based on relaxation of the nuclear or possibly even the electronic spins. Again such fits are relatively good for temperatures below T = 70 K (not shown) though the $\chi^{2}$ = 1.44 $\pm$ 0.14  is slightly worse. Again the width of the central Lorentzian becomes narrower with increasing temperature (from a FWHM of 1.84 $\pm$ 0.18 \ $\mu$eV at 2 K to a FWHM of 1.22 $\pm$ 0.12 \ $\mu$eV at 80 K) and its fitted intensity disappears above 85 K. However, as this model is based on relaxations, a narrowing of the width with temperature would mean a slowing down of the relaxations with increasing temperature which we judge to be unphysical.

\begin{figure}
\resizebox{0.45\textwidth}{!}{\includegraphics{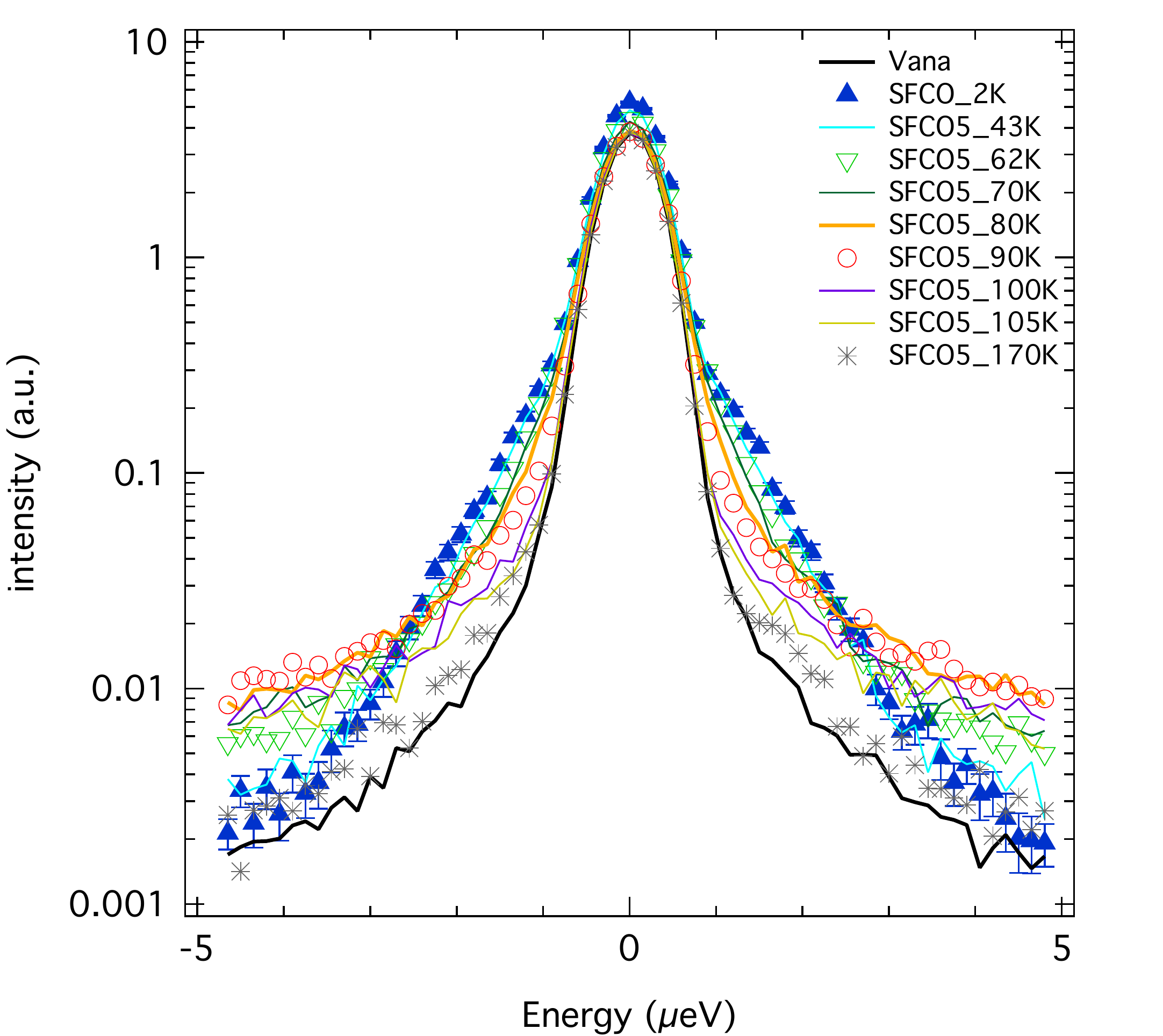}}
\caption {Temperature dependence of the Sr$_2$FeCoO$_6$ (SFCO) spectra at 2 K compared to the Vanadium resolution (summed from Q = 0.65 {\AA}$^{-1}$ to  1.9 {\AA}$^{-1}$). The legend shows the temperatures as a suffix. Due to a factor 10 improved signal-to-noise ratio of the IN16B HSNR mode and the energy range extending to 5 $\mu$eV compared to Fig.\ref{SFCO_Tdep_pol} the presence of additional quasielastic scattering between T = 62 K and 170 K becomes clear.}
 \label{compareHSNR5_SLFCO_SFCO_allTemp}
\end{figure}

We have then re-measured Sr$_2$FeCoO$_6$ spectra at some temperatures in the HSNR mode of IN16B and have extended as well the energy transfer range to $\pm$5 and $\pm$30 $\mu$eV, respectively, with the aim to search for possible quasielastic scattering. A comparison of Sr$_2$FeCoO$_6$ and SrLaFeCoO$_6$ spectra with the Vanadium resolution function in Fig.\ref{SLFCO_SFCO_5ueV_HSNR} evidences clearly that we observe for both samples hyperfine splitting.

The temperature dependent spectra for Sr$_2$FeCoO$_6$ measured in HSNR mode are shown in Fig.\ref{compareHSNR5_SLFCO_SFCO_allTemp}. We see at energies below 3 $\mu$eV a narrowing of the spectra, consistent with the high resolution measurements above. However, further out in energy transfer in the range from 3 - 5 $\mu$eV this narrowing goes along with an increase of the spectral intensity, which reaches its maximum near the transition temperature at T $\approx$ 80 - 90 K before decreasing again to the level of the low temperature spectrum. This shows clearly the appearance of quasielastic scattering similar to SrLaFeCoO$_6$ in Fig.\ref{integrate_inelastic_Intensity}, but with the important difference that quasielastic scattering is now observed as well at high Q. The energy window of $\pm$ 3 $\mu$eV chosen previously with the polished crystal setup was too narrow and the signal-to-noise ratio insufficient to detect this intensity increase. Both measurements together, however, evidence the existence of hyperfine splitting which softens near the transition temperature simultaneously with the uprising of additional quasielastic scattering which broadens with temperature. 

Applying a fit to the spectrum measured with HSNR and limiting it to the lowest temperature where no quasielastic scattering is observed, interestingly the low temperature spectrum of Sr$_2$FeCoO$_6$ can also be quite well fitted with two different sets of hyperfine splitting triplets ($\chi^{2}$  = 0.76), an additional elastic contribution and a flat background. The first more intense hfs spectrum (blue dotted-dashed fit component) has an energy splitting of only $\Delta$E$_1$ = 0.91 $\pm$ 0.018 $\mu$eV, which leads to a plateau like curve and to an inelastic shoulder in the spectra. The second hfs spectrum is by a factor 10 weaker and has an energy splitting of $\Delta$E$_2$ = 1.98 $\pm$ 0.044 $\mu$eV. All hyperfine split - lines were constrained to the same width and are with FWHM $\approx$ 0.52 $\mu$eV slightly broadened. At T = 43 K the hfs splittings would be  0.736 $\pm$ 0.062 $\mu$eV and 1.83 $\pm$ 0.13 $\mu$eV ($\chi^{2}$  = 1.13). At higher temperature, near T = 62 K such fits start to become inconsistent, showing a hyperfine splitting of 0.20 $\pm$ 0.08 and 2.23 $\pm$ 0.1($\chi^{2}$  = 1.66), at T = 70 K  0.20 $\pm$ 0.033 $\mu$eV and 2.02 $\pm$ 0.05 $\mu$eV ($\chi^{2}$  = 1.08), at T = 80 K  0.20 $\pm$ 0.08 and 2.035 $\pm$ 0.077 $\mu$eV ($\chi^{2}$  = 1.38) and at T = 90 K and higher the intensity of these fit components becomes negligible and $\chi^{2}$ continues to increase.

\begin{figure}
\resizebox{0.45\textwidth}{!}{\includegraphics{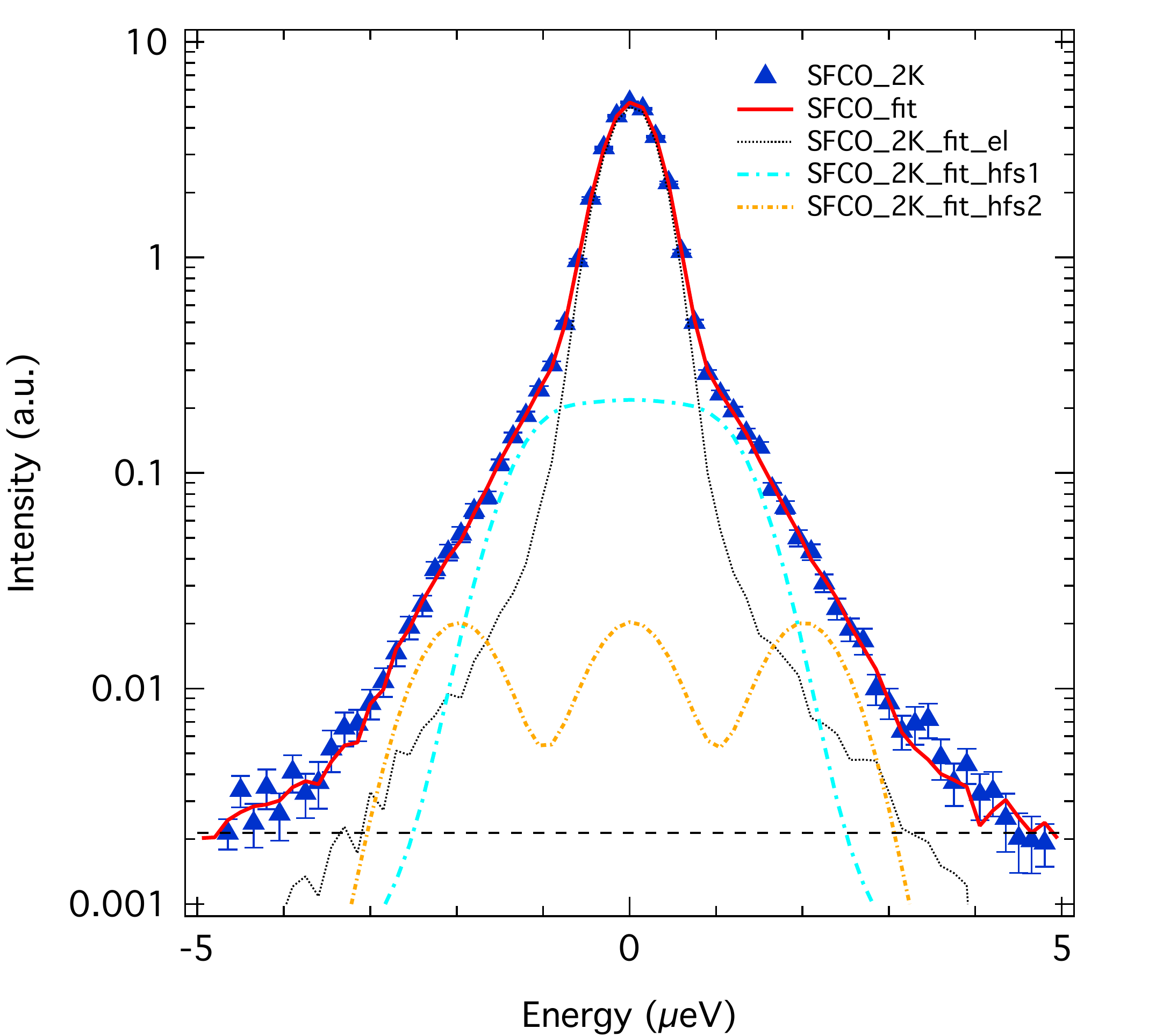}}
\caption {Fit of the Sr$_2$FeCoO$_6$ spectrum at T = 2 K with a hyperfine spectrum, additional elastic component and quasielastic component fitted by a Lorentzian.}
 \label{fit_SFCO_2K_El2hfsBg}
\end{figure}

It is interesting to note that the larger splitting is very close to the Co - hfs observed for SrLaFeCoO$_6$ and therefore it is tempting to speculate that this spectrum may result from two distinct local environments around the Co - atoms and we may speculate that this might be related to the B-site disorder mentioned in the beginning of this chapter.

\begin{figure}
\resizebox{0.45\textwidth}{!}{\includegraphics{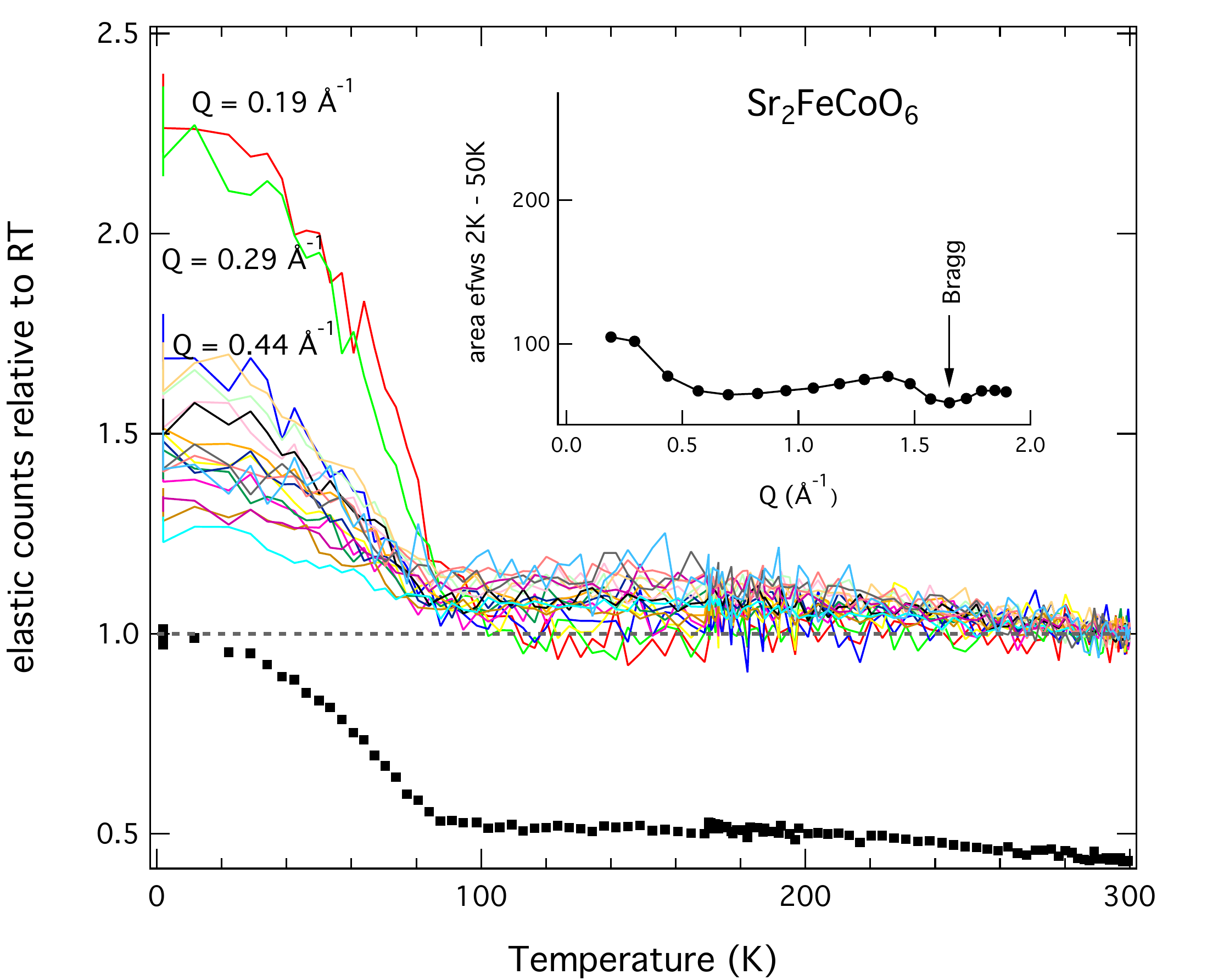}}
\caption {Temperature-dependence of the elastic intensity of Sr$_2$FeCoO$_6$ normalised to its high temperature average value. for all Q values measured. Main panel, bottom (black squares): sum over all Q-values; for clarity offset by -0.57. Lines in main panel: elastic intensity for different Q values.  A strong increase  of the elastic intensity is observed below T = 80 K.  Inset: The Q -dependence of the area under the efws curves between T = 2 and 50 K. Besides the increase at low Q, a small increase of an elastic diffuse contribution can also be observed around the Bragg peak position at Q = 1.65 {\AA}$^{-1}$.}
 \label{SFCOefws}
\end{figure}

Finally investigating again temperature and Q-dependence of the elastic intensity for Sr$_2$FeCoO$_6$ (Fig. \ref{SFCOefws}), like for SrLaFeCoO$_6$ discussed above, we find in cooling qualitatively the same behaviour with a slow intensity increase between 300K and 100K and a steep increase near T$_{sf}$ towards low temperatures. The Q-dependence for the elastic intensity raise at low temperature is again most pronounced for lowest Q-values and again we observe a diffuse peak below the high temperature Bragg peak position(inset to Fig. \ref{SFCOefws}).

\subsection{Magnetically ordered  La$_2$FeCoO$_6$}

\begin{figure}
\resizebox{0.45\textwidth}{!}{\includegraphics{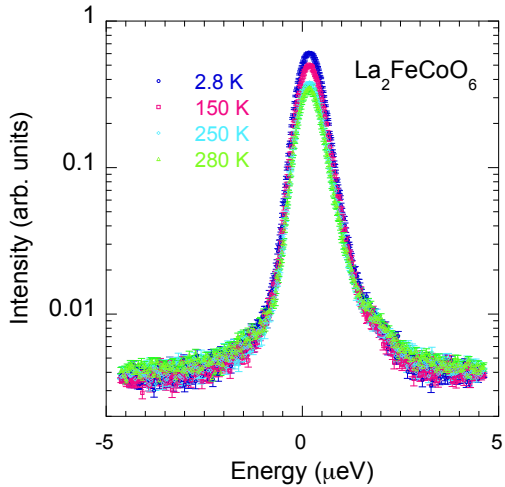}}

\caption {Spectra of La$_2$FeCoO$_6$ at different temperatures. There exist no extra scattering at low temperatures.
 }
 \label{LFCO-spectra}
\end{figure}

\begin{figure}
\resizebox{0.45\textwidth}{!}{\includegraphics{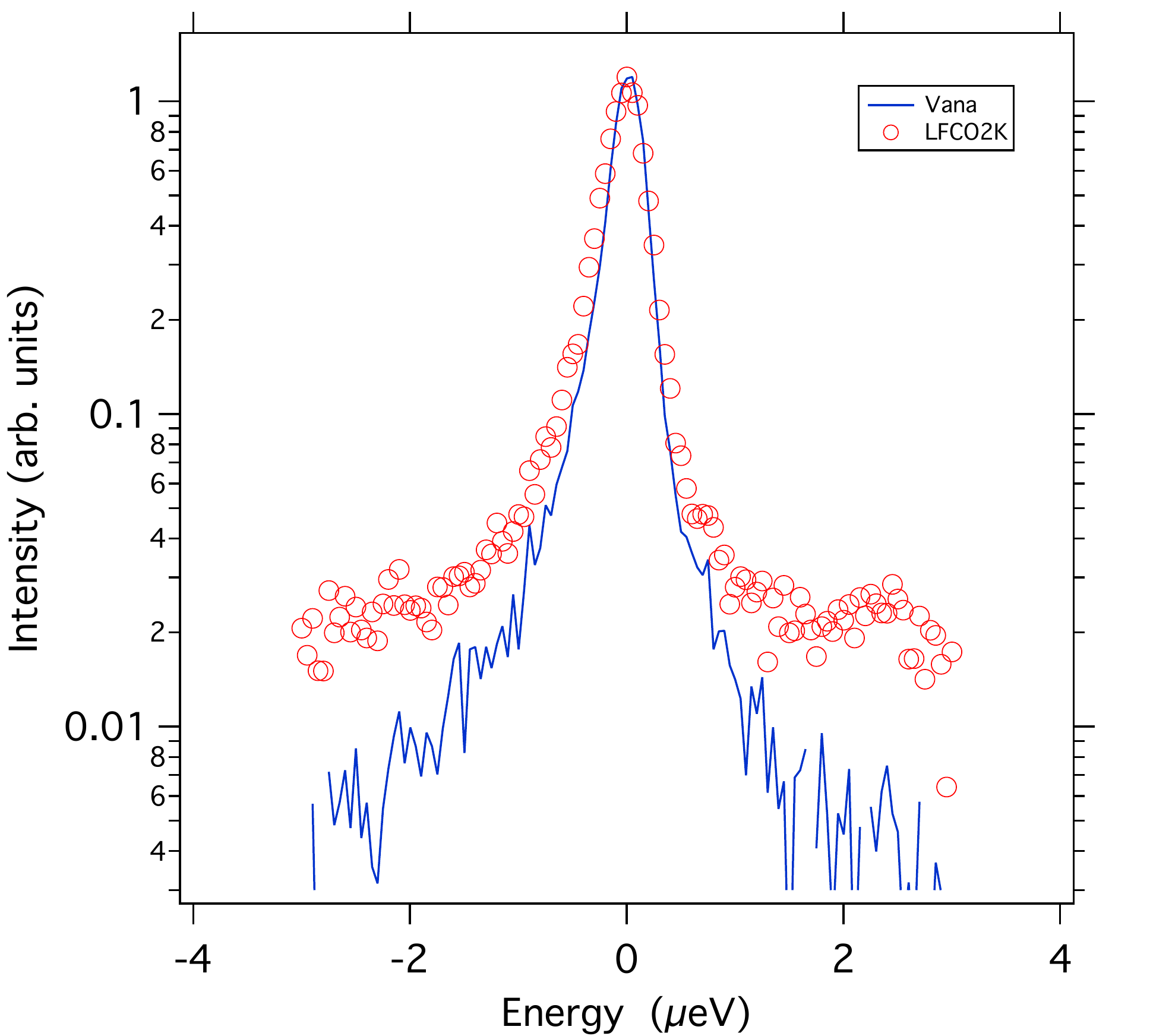}}
\caption {Spectra of La$_2$FeCoO$_6$ measured on IN16B with higher energy resolution. The background corrected data are scaled to peak maximum.}
 \label{LFCO_compareVana}
\end{figure}

\begin{figure}
\resizebox{0.5\textwidth}{!}{\includegraphics{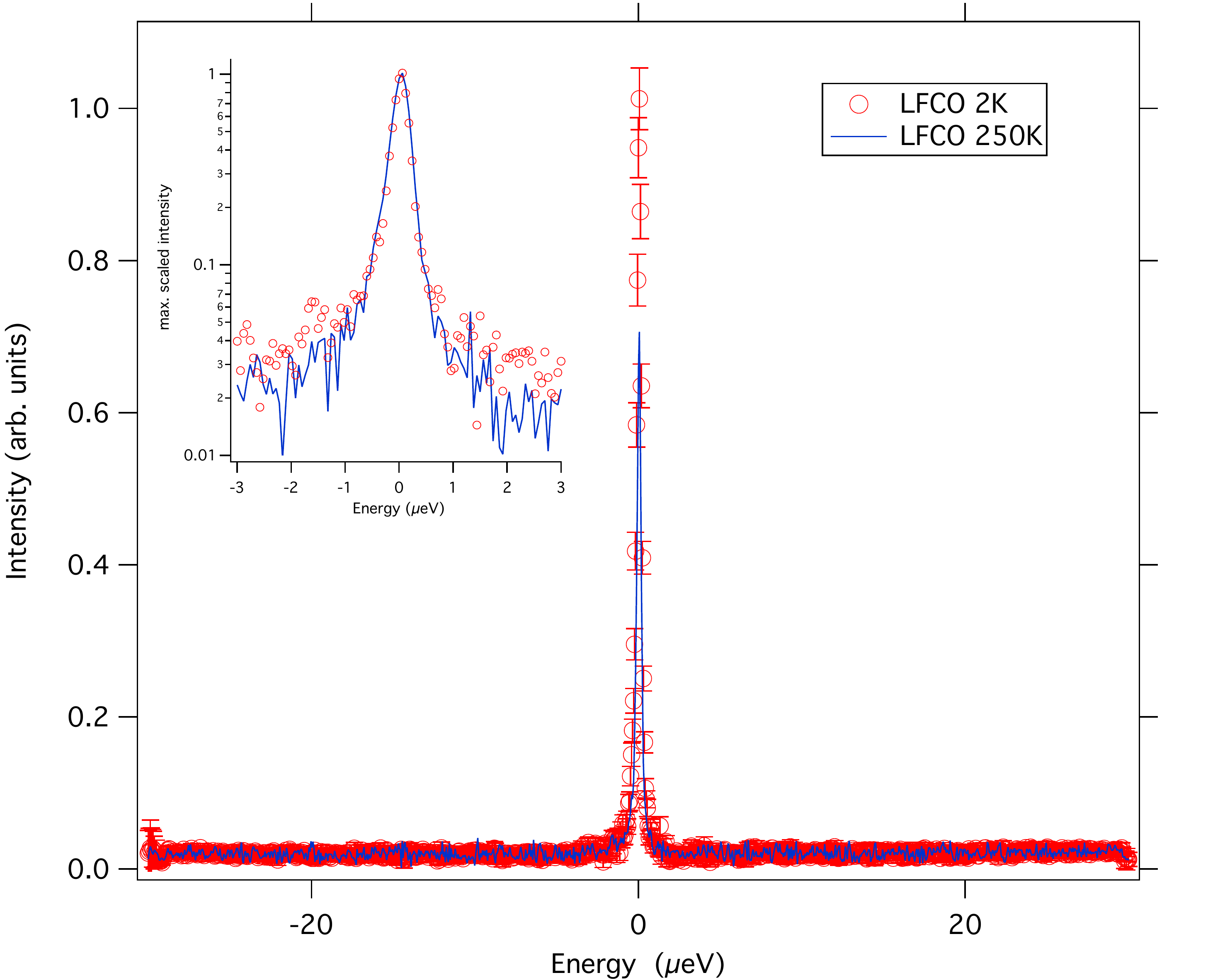}}
\caption {Spectra of La$_2$FeCoO$_6$ measured on IN16B with higher resolution and over a wider energy transfer range. The data are plotted on a linear intensity scale and are not normalised. The inset shows the near elastic region on a logarithmic intensity axis scaled to peak maximum.
 }
 \label{LFCO_30ueV}
\end{figure}

\begin{figure}
\resizebox{0.5\textwidth}{!}{\includegraphics{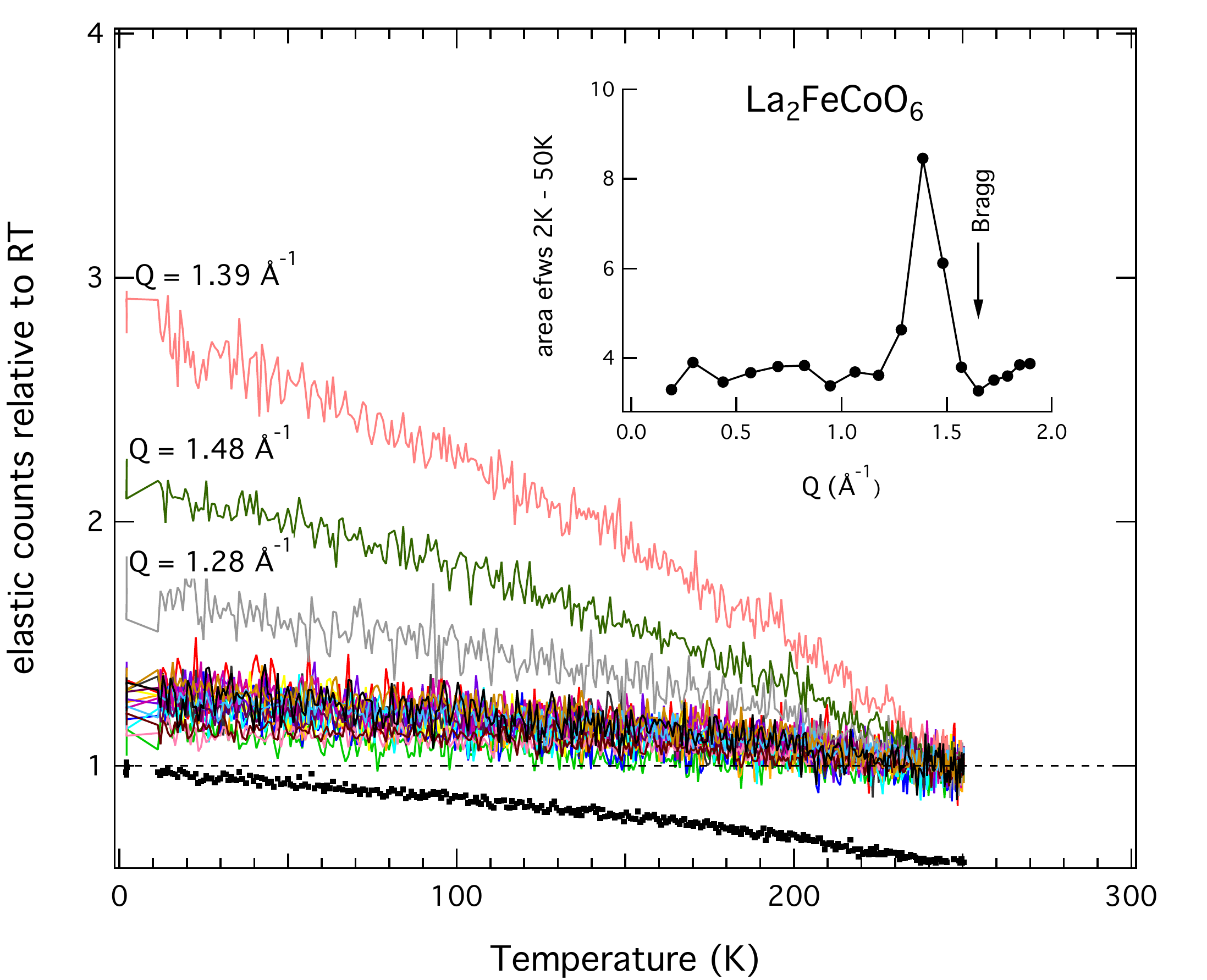}}
\caption {Temperature-dependence of the elastic intensity of La$_2$FeCoO$_6$ normalized to its high temperature average value. Main panel, bottom (black squares): sum over all Q-values; for clarity offset by -0.6. Lines in the main panel: elastic intensity for different Q values, with the three Q-values of the low temperature Bragg peak labelled. Besides at the additional Bragg peak the elastic intensity increases only weakly with decreasing temperature. Inset: The Q -dependence of the area under the efws curves between T = 2 and 50 K. No additional increase at low Q is observed, only the appearance of an additional Bragg peak at Q$\approx$1.4{\AA}$^{-1}$ below the high temperature Bragg peak position marked with an arrow at Q = 1.61 {\AA}$^{-1}$. }
 \label{LFCOefws}
\end{figure}

High-resolution neutron diffraction study showed that  La$_2$FeCoO$_6$ orders \cite{chatterji} magnetically below about 225 K, and there is also a structural transition at the same temperature. Simulation work predicts a ferromagnetic semiconductor state \cite{fuh, labrim}without specifying a transition temperature . 

However, to our surprise in spite of this magnetic ordered state we observe for La$_2$FeCoO$_6$ down to 1.8 K no measurable quasi- or inelastic signal on both backscattering spectrometers which could be a hint for hfs. The spectra taken on SPHERES at four different temperatures are shown in Fig.\ref{LFCO-spectra} and additional measurements on IN16B using the high resolution configuration comparing either the spectra at 2K with Vanadium (Fig.\ref{LFCO_compareVana}) or with the 250K spectrum (Fig.\ref{LFCO_30ueV}) confirm this observation. On IN16B the Doppler drive was set both for an energy transfer of $\pm$ 3 \ $\mu$eV (Fig.\ref{LFCO_compareVana}) and additionally to $\pm$ 30 \ $\mu$eV (Fig.\ref{LFCO_30ueV}), but no sign of hyperfine signal was detected in the whole energy range. Though from Fig.\ref{LFCO_compareVana} one might get the impression that the elastic line could be broadened,  fits with a resolution (Vanadium) convoluted delta function plus a zero centred Gaussian do not show a significant broadening. This is also consistent with the measurement extended over $\pm$ 30 \ $\mu$eV in Fig.\ref{LFCO_30ueV} showing that only the elastic intensity has decreased between 2 K and 250 K.

Thus data from both instruments indicate only that there might be a background increase with temperature, but no sign of an inelastic hyperfine splitting. The implication for this negative results is obviously that the hyperfine field at the Co nuclear site is too weak and that Co-ions magnetic moments may not get ordered or frozen at low temperatures. But since we probe only locally the hfs on the Cobalt site we can not exclude an ordering of Fe spins. 

For La$_2$FeCoO$_6$ we have again analysed on IN16B the elastic fixed window scans (efws) as function of temperature for which raw data as a function of Q and T are shown in the Appendix, Fig.\ref{LFCO_efws_3D} as a 3D-plot. In cooling we see clearly how an additional Bragg peak emerges in the high Q-range below the Bragg peak position of the high temperature phase (Q$\approx$1.65{\AA}$^{-1}$) though we cannot say if its origin is nuclear or magnetic.  
Again we compare in Fig.\ref{LFCOefws} the relative elastic intensity increase in cooling relative to 250K. In contrast to the first two samples discussed, La$_2$FeCoO$_6$ orders at high temperature ($T \approx 225$ K) and the elastic intensity increases only gradually without a clear step, as shown for the Q-averaged intensity in the lower part of the main panel (black dots) and for the individual Q-values (coloured lines) in Fig \ref{LFCOefws}. Whereas our scans do not start sufficiently far above the supposed transition temperature  T $\approx$ 225 K to exclude a potential step like change above, nearly identical spectra observed at 250 K and 280 K on SPHERES (Fig.\ref{LFCO-spectra}) render this highly improbable. It should be stressed that in Fig.\ref{LFCOefws} the curves with a stronger relative elastic intensity increase in cooling (labeled with their respective Q-values) belong here to the high Q-range where an additional Bragg peak was found, near Q$\approx$1.4{\AA}$^{-1}$. This is clearly seen from the inset to this figure, displaying the Q-dependence of the area under the efws from T = 2 and 50 K.

\section{Discussion}

We have presented results from high energy resolution neutron backscattering on three double perovskite samples of type A$_2$BB$^{\prime}$O$_6$ with B:B$^{\prime}$=1:1=Fe:Co where the occupation of the A-site was either complete with La and Sr, respectively, or 1:1 with La:Sr. The three investigated samples have completely different magnetic ground state, which therefore can not be ascribed to the unchanged B-site disorder alone. From literature it is known that at low temperature Sr$_2$FeCoO$_6$ is a canonical spin glass state, SrLaFeCoO$_6$ a magnetic glass and La$_2$FeCoO$_6$ magnetically ordered. Our inelastic neutron scattering measurements serve first as a local probe for the hyperfine field which might or not be induced at the Cobalt nucleus by the electronic spins. Second, the additionally investigated temperature dependence of the efws reveals changes which might stem from the electronic magnetic moments of both the Co and Fe ions, for which besides cation site disorder a valence state disorder seems to be of importance. A slowing down of electronic spin fluctuation, e.g. by spin freezing at low temperature, should show up in the efws as an intensity increase, whereas the appearance of hfs at low temperatures should be reflected rather in an elastic intensity decrease on cooling as the inelastic spin flip scattering separates from the elastic line. The efws could in principle be further influenced by structural changes (coherent diffraction contributions) or atomic diffusion (mainly incoherent, but also coherent).

\begin{figure}
\resizebox{0.45\textwidth}{!}{\includegraphics{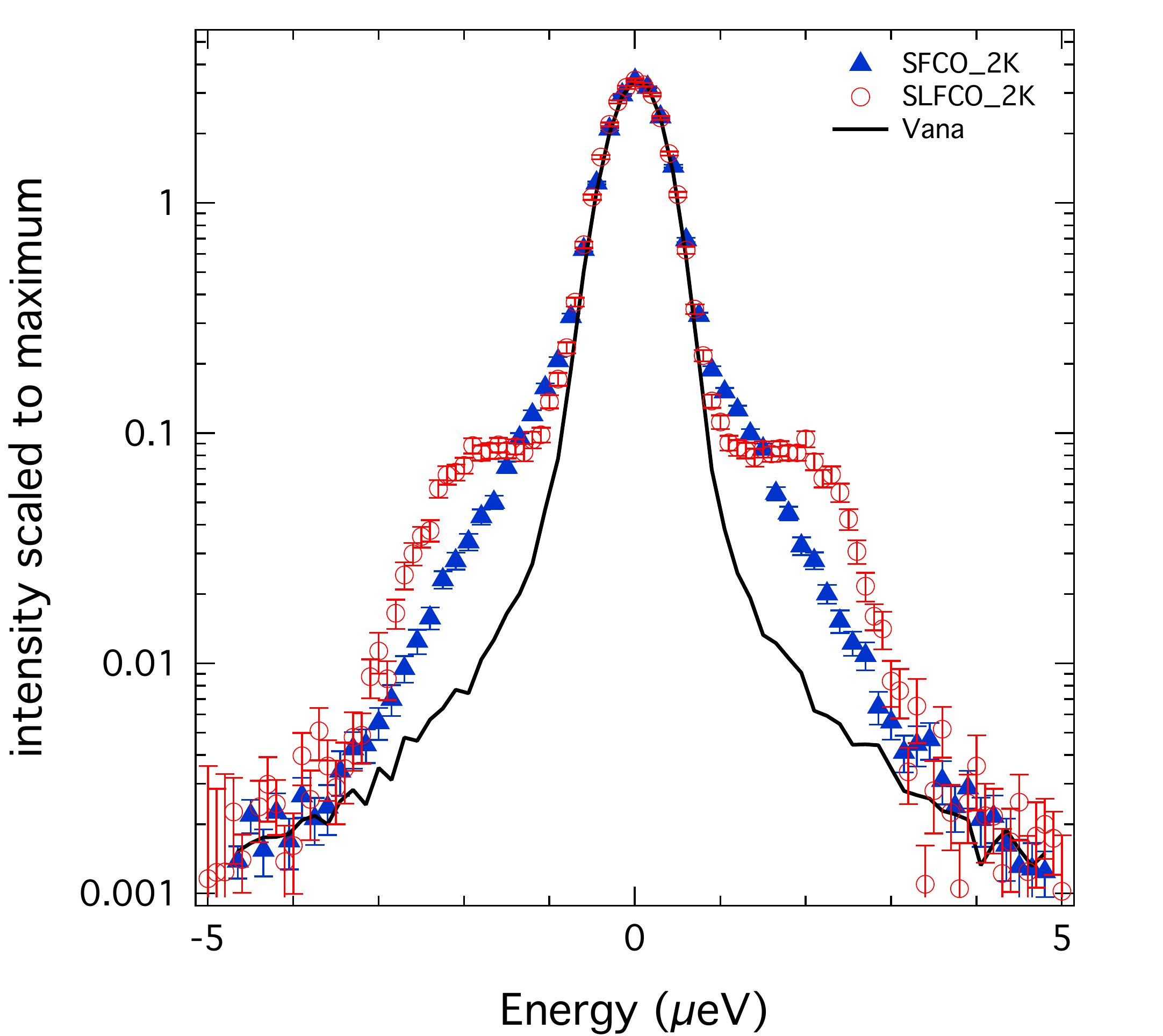}}
\caption {Comparison of SrLaFeCoO$_6$ (red open circles) and Sr$_2$FeCoO$_6$ (blue filled triangles) spectra measured at T = 2 K compared to the Vanadium resolution spectrum (line). All spectra are background corrected and summed between Q = 0.44 and 1.9 {\AA}$^{-1}$. The intensities are normalised to the elastic peak maximum. }
 \label{SLFCO_SFCO_5ueV_HSNR}
\end{figure}

As for the inelastic measurements we observe for both Sr$_2$FeCoO$_6$ and SrLaFeCoO$_6$ extra scattering outside of the resolution which can be ascribed to the existence of hyperfine splitting at the Co-site. This is summarised for for all samples at T = 2 K in Fig.\ref{SLFCO_SFCO_5ueV_HSNR}. SrLaFeCoO$_6$ shows clear inelastic scattering as known for hfs and model fits show broadened peaks which we ascribe to disorder in the electronic spin system. The spectra for Sr$_2$FeCoO$_6$ are significantly narrower, probably due to a weaker local field at the Co-nucleus. Finally for La$_2$FeCoO$_6$ (Fig. \ref{LFCO_compareVana}) we find no extra inelastic scattering within the applied instrumental resolution range. The temperature dependence of the hfs spectra for both the spin glass, Sr$_2$FeCoO$_6$, and the magnetic glass, SrLaFeCoO$_6$, resemble the conventional hfs behaviour of other systems, like the softening of the inelastic peak positions and a final merging with the elastic line (see Appendix, Figures: \ref{SLFCO-ET},\ref{SFCO_hfs_fits_PosInt}). In contrast, hardly any change of the inelastic spectra is observed with temperature for the magnetically ordered La$_2$FeCoO$_6$.  

\begin{figure}
\resizebox{0.45\textwidth}{!}{\includegraphics{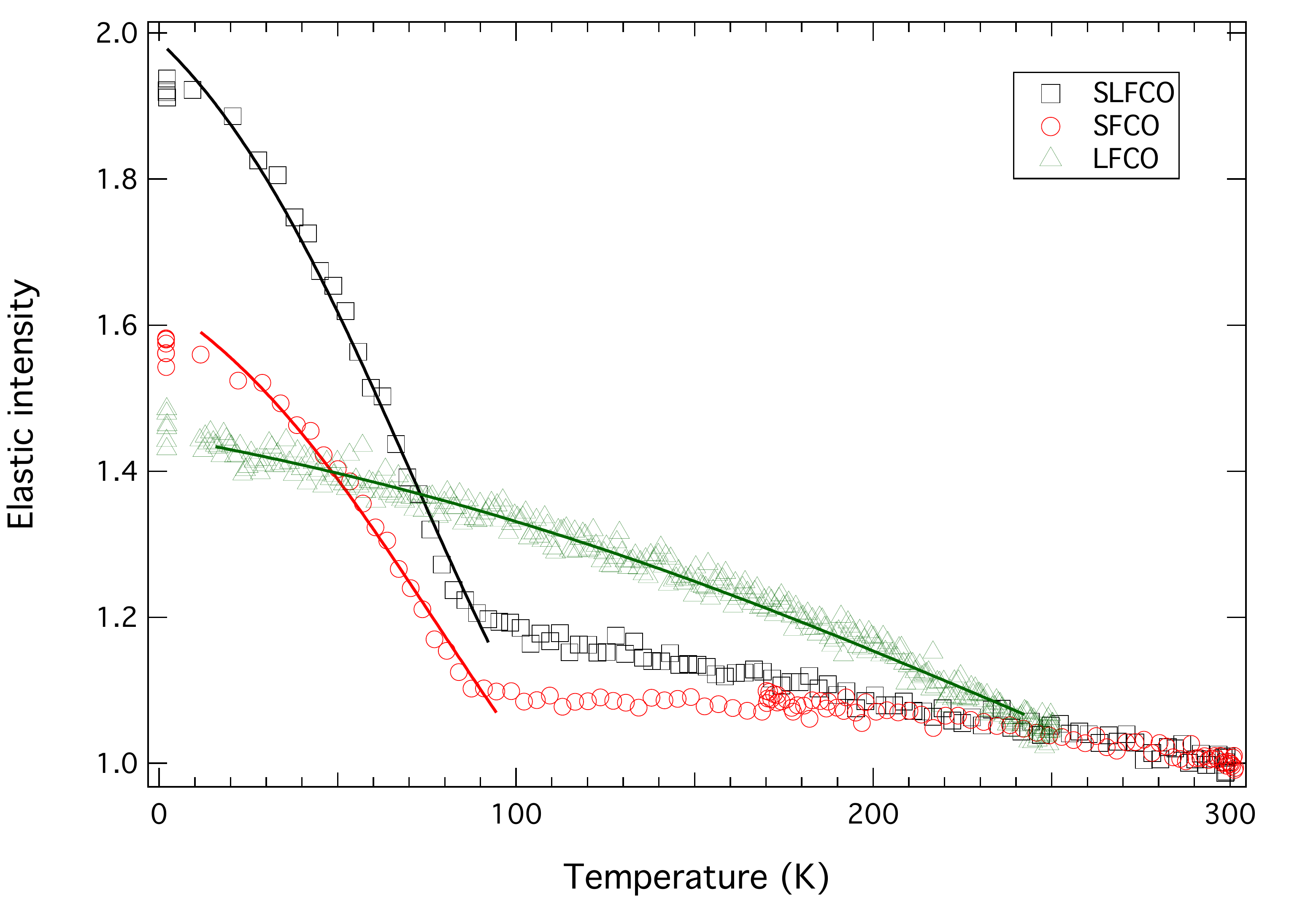}}
\caption {Elastic intensity of La$_2$FeCoO$_6$ (green triangles) measured on IN16B in high resolution mode summed over all detectors, compared to the elastic intensity of Sr$_2$FeCoO$_6$ (red circles) and SrLaFeCoO$_6$ (black squares). The data are normalised to the highest temperatures measured and for La$_2$FeCoO$_6$ to the T=250K value of the others. For comparison La$_2$FeCoO$_6$ is scaled to the 250K value of the other curves. Lines are fits with Brillouin functions with Tc fixed to 75K and 250K.}
 \label{elas_compareAll}
\end{figure}

The temperature dependence of the elastic intensity for the investigated samples is summarised in Fig.\ref{elas_compareAll}. For the relative elastic intensity of SrLaFeCoO$_6$ and Sr$_2$FeCoO$_6$, which are magnetically disordered, but showed signatures of hyperfine splitting in the inelastic neutron spectra, we find in the efws a clear step-like increase near the spin freezing temperatures known from literature. For the efws of the magnetically ordered La$_2$FeCoO$_6$ we observe in cooling only a smooth elastic intensity increase. The onset temperatures of the step-like increase of the elastic intensity for SrLaFeCoO$_6$ and Sr$_2$FeCoO$_6$ and the temperatures where the hfs becomes visible coincide roughly. As the appearance of hfs would rather lead to a decrease of the efws near the transition temperature we postulate that the observed elastic intensity increase is due to an ordering or freezing of the electronic spins, which is at the origin of the hfs at the Co-site. That the relative increase of the efws step is smaller for Sr$_2$FeCoO$_6$ compared to SrLaFeCoO$_6$ is consistent with an observed smaller hfs or smaller local field at the Co-site for Sr$_2$FeCoO$_6$. 

The Q-dependence of the efws gives other useful information which can be compared to the known diffraction results. For the ordered magnetic system, La$_2$FeCoO$_6$, we find in the elastic scattering the known Bragg peaks, especially we find the additional Bragg peak appearing below T $\approx$ 225 K near Q$\approx$1.4{\AA}$^{-1}$. For the disordered magnetic systems, however, no long range or short range magnetic order has been reported. We find at lowest temperatures, where the electronic spins are assumed to be frozen, in the Q-dependence of the efws for both magnetically disordered systems diffuse scattering near Q$\approx$1.4{\AA}$^{-1}$, the region where for the magnetically ordered La$_2$FeCoO$_6$ a Bragg peak appears. Furthermore both, SrLaFeCoO$_6$ and Sr$_2$FeCoO$_6$, show an increase of the elastic intensity at small Q, which is again most pronounced for SrLaFeCoO$_6$ and not present for La$_2$FeCoO$_6$.

At last we point out that quasielastic scattering appears near the reported freezing temperatures which suggests spin fluctuations on the ns-time-scale. This quasielastic scattering is visible only at small Q for the magnetic glass SrLaFeCoO$_6$, it is visible as well at higher Q for the spin glass Sr$_2$FeCoO$_6$ and either missing or too wide for the ordered La$_2$FeCoO$_6$. Thus a possible speculation could be that with the addition of Sr in La$_{2-x}$Sr$_x$FeCoO$_6$ the spin fluctuations slow down. In La$_2$FeCoO$_6$ the spins could fluctuate very fast with the corresponding quasielastic signal being wider than the energy range of IN16B. In the magnetic glass the fluctuations would be relatively fast too and only be visible in the low Q-, long-range region, maybe related to postulated magnetic domains in this system. Finally for the spin glass Sr$_2$FeCoO$_6$ the fluctuations would be more localised and further slowed down, so that quasielastic scattering becomes visible even for higher Q. At the same time the complete freezing of spins, related to the relative step height of the efws, would become more rare. We are aware that our data give only weak support for such speculations and that the real explanation might be much more complex, but at least they sketch some trend for an otherwise non-monotonous behaviour of the measured hfs with he La/Sr-ratio.

\section{Conclusion}

We have presented the first results on the hyperfine field of Cobalt obtained by inelastic neutron scattering for the double perovskites La$_{2-x}$Sr$_x$FeCoO$_6$ for x=0,1,2. An important conclusion from these experiments on Sr$_2$FeCoO$_6$, SrLaFeCoO$_6$ and La$_2$FeCoO$_6$ is that the first two are magnetically disordered systems and still show hyperfine splitting, whereas for the magnetically ordered La$_2$FeCoO$_6$ system no hf-splitting is observed. This observation does not conform with changes controlled in proportion to the Sr concentration x, and indicates that the effect of A-site doping in AA$^\prime$FeCoO$_6$ are more complex. The same is valid in what concerns the temperature dependence of the elastic intensity, for which a step like increase at low temperatures is interpreted as freezing of the electronic spin fluctuations. These measurement may incite further studies on this interesting class of double perovskites.

\section{Appendix}

\subsection{SrLaFeCoO$_6$}

\begin{figure}
\resizebox{0.45\textwidth}{!}{\includegraphics{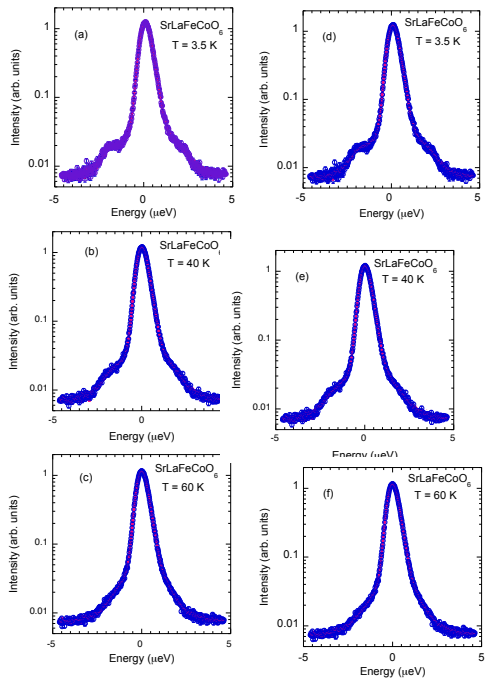}}
\caption { Typical fit of the spectra from SrLaFeCoO$_6$ measured on SPHERES by delta function convoluted with the resolution function for the elastic peak and for the inelastic peaks on the energy loss and the energy gain sides were Gaussian functions convoluted with resolution function. We essentially used two types of fitting procedures. In the first procedure we kept the widths same but let it vary (a,b,c) whereas in the second procedure we kept the widths fixed to the low temperature value determined (d,e,f).}
 \label{SLFCO-fit}
\end{figure}

\begin{figure}
\resizebox{0.45\textwidth}{!}{\includegraphics{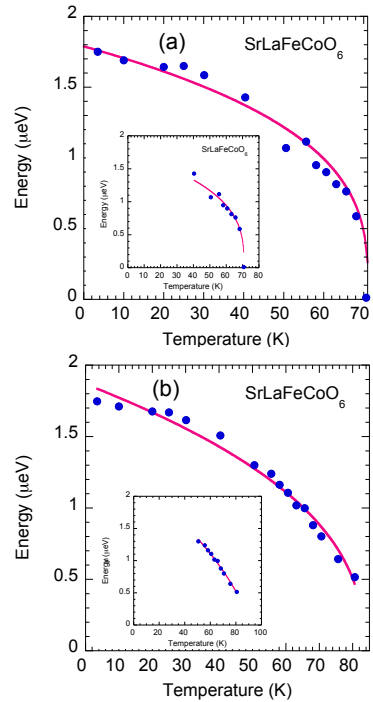}}
\caption {Power-law fit to the temperature dependence of the energy of hyperfine splitting in SrLaFeCoO$_6$: (a) fit in which the widths of the inelastic peaks are constrained to be equal but was varied and (b) fit in which equal widths of the inelastic peaks were fixed to the low temperature value. The fits have been shown for all data in the main figure whereas the insets show the fits only with data measured at temperatures above T = 40 K only. The magnetic transition temperature obtained by these fits are are appreciably different.
 }
 \label{SLFCO-ET}
\end{figure}

Fit details for the IN16B spectra were reported above. In what concerns fitting of the SPHERES spectra (Fig.\ref{SLFCO-fit}) we have essentially used two different procedures. In the first procedure we constrained the widths of two inelastic peaks at the energy loss and the energy gain sides equal but let it vary whereas in the second procedure we kept the widths same and fixed to the low temperature value determined. These two procedures yield different energies for the hyperfine splitting. The resulting average energy of hyperfine splitting as a function of temperature is shown in Fig. \ref{SLFCO-ET} along with power law fits for the two different procedures. We have shown previously that in ordered magnetic systems \cite {chatterji09} the the energy of hyoperfine splitting is the order parameter of the magnetic phase transition. So one could fit power-law exponent to extract the critical exponent in favorable cases. Here we don't have an magnetic ordering but spin freezing. Also we do not have many data points close to the transition temperature; so a power-law fit can perhaps be used only to parametrize the temperature dependence of the hfs. We have done this here by using   
\begin{equation}
E(T) = A\left(\frac{T_{sf} -T}{T_{sf}}\right)^{\beta} 
\end{equation}
 where $E$ is the energy splitting,  $T_{sf}$ is the magnetic transition or spin-freezing temperature $T_{sf}$ and $\beta$ is the power-law exponent. We note that two different procedures give different values of $T_{sf}$ and $\beta$. The transition temperature obtained by varying the line width gave $T_{sf} = 71.2 \pm 0.8$ K and $\beta = 0.32 \pm 0.05$, whereas the transition temperature obtained by fixing the peak widths gave $T_{sf} = 83 \pm 1$ K and $\beta = 0.41 \pm0.03$. Normally the power-law fit should not hold for the entire low temperature range and we have also fitted the power-law only with data for temperatures greater than about 40 K. These fits are given in the insets of the figures and the corresponding fit results were : $T_{sf} = 70.8 \pm 0.3$, $\beta = 0.33 \pm 0.07$ and  $T_{sf} = 90 \pm 3$, $\beta = 0.7 \pm 0.1$. We mote that the $T_{sf}$  obtained by two different fit procedures are also different by more than 10 K. 

\begin{figure}
\resizebox{0.45\textwidth}{!}{\includegraphics{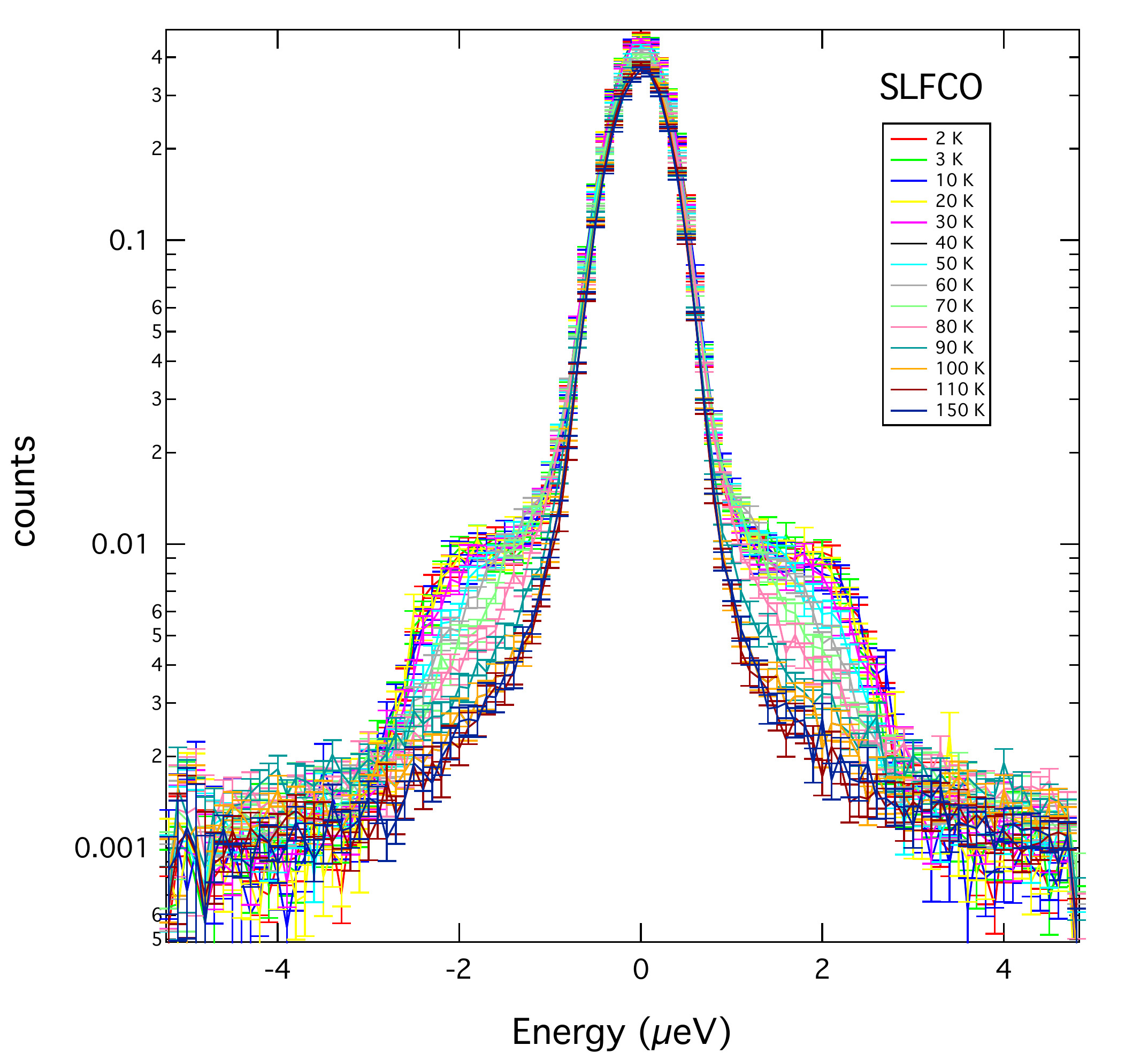}}
\caption {Spectra for SrLaFeCoO$_6$ measured on IN16B during cooling in the standard high flux mode; sum over Q above 0.44 {\AA}$^{-1}$ are shown.
 }
 \label{SLFCOspectra_HFnormal_TdepSums}
\end{figure}

SLFCO was first measured during a test at IN16B in the normal HF configuration (see Fig.\ref{SLFCOspectra_HFnormal_TdepSums}). Spectra were measured during cooling. The data for Q-dependent elastic and inelastic intensities in Fig.\ref{integrate_inelastic_Intensity} were deduced from these spectra. The inelastic peaks due to hfs and the merging to the elastic with increasing temperature can be seen.

\subsection{Sr$_2$FeCoO$_6$}

\begin{figure}
\resizebox{0.35\textwidth}{!}{\includegraphics{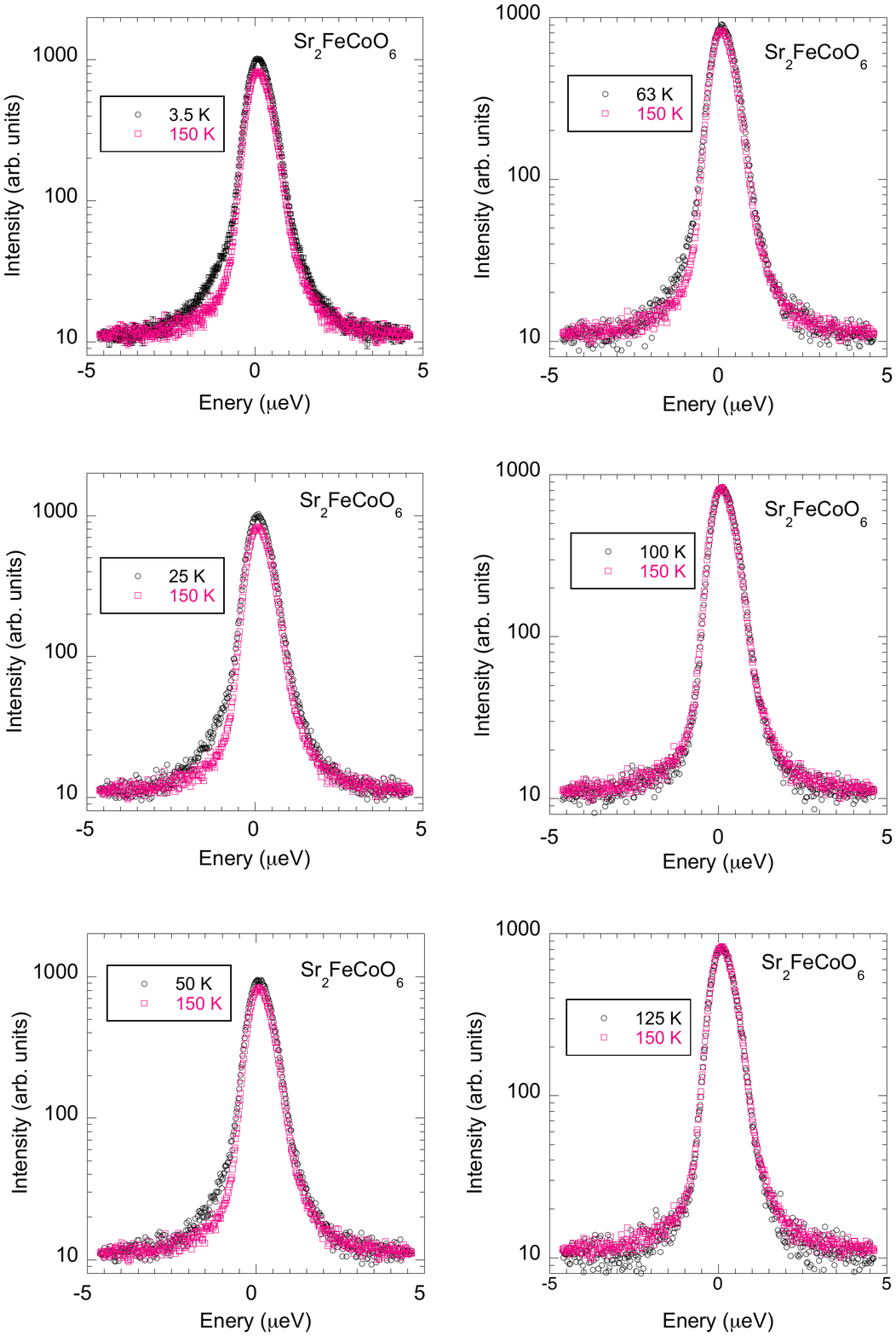}}
\caption {Spectra of Sr$_2$FeCoO$_6$ measured on SPHERES at different temperatures compared with the spectrum at T = 150 K. There exist extra scattering at low temperatures below the spin temperature $T_{sf} \approx 80$ K.
 }
 \label{SFCO}
\end{figure}

\begin{figure}
\resizebox{0.3\textwidth}{!}{\includegraphics{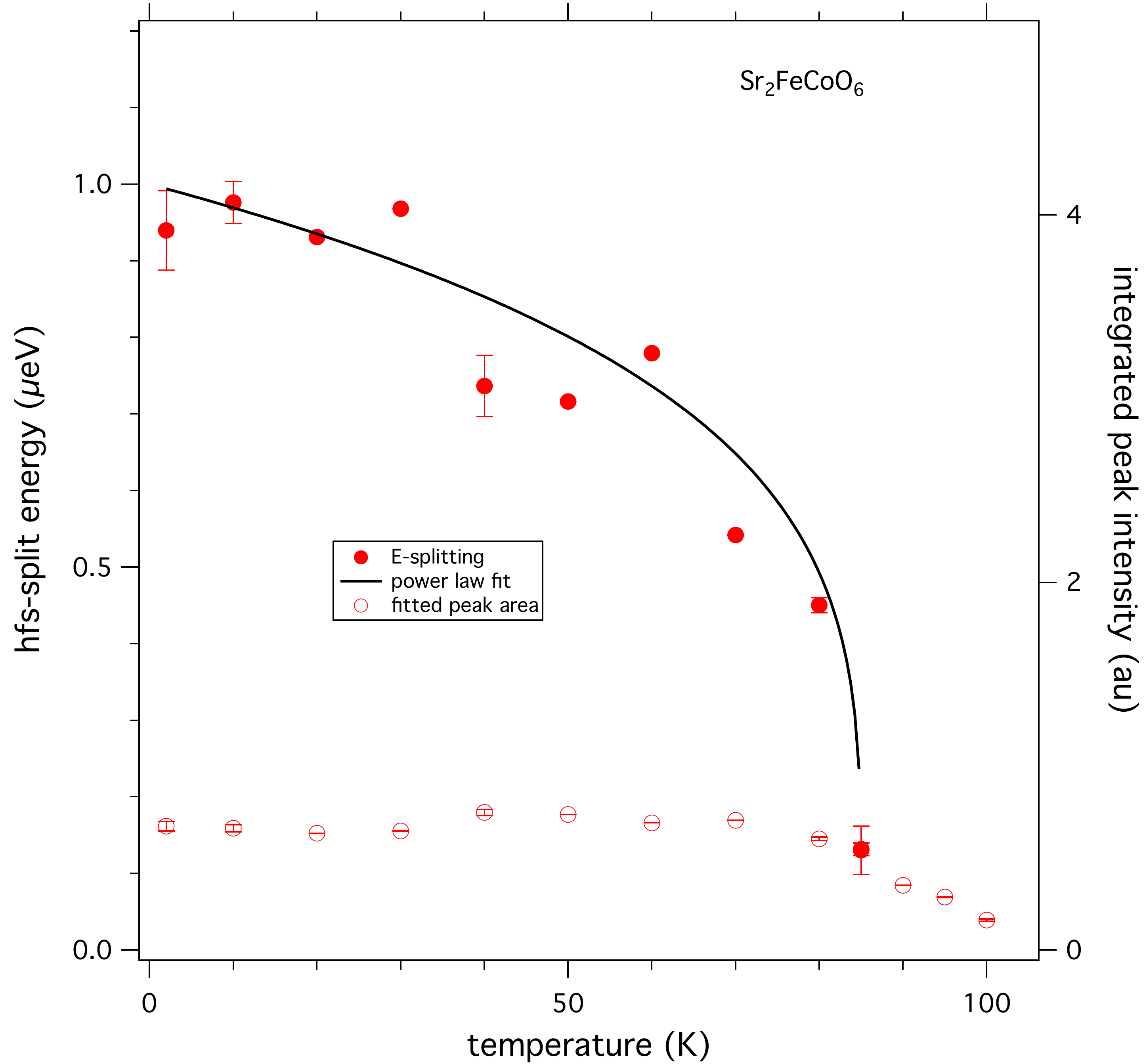}}
\caption {Temperature dependence of the peak positions of Sr$_2$FeCoO$_6$ as result from fitting the IN16B spectra with model A. The temperature dependence was arbitrarily fitted with a power law (solid line) with exponent 0.25 and a critical temperature of T = 85 K at which the area under the peak triplet (open circles) disappears.}
 \label{SFCO_hfs_fits_PosInt}
 \end{figure}

Fig.\ref{SFCO} shows the spectra for Sr$_2$FeCoO$_6$ measured at SPHERES at different temperatures compared with the spectrum at T = 150 K. The spectrum at T = 3.5 K has clearly extra signal compared to that at T = 150 K, which we considered as the resolution function. However the resolution at SPHERES was not sufficient to fit the hyperfine splitting. Above, in the main paper, we have shown fit curves with different models for IN16B spectra, measured on Sr$_2$FeCoO$_6$ at 2K with higher resolution and/or better signal-to-noise ratio. Here we summarise the fit parameters: In Fig. \ref{SFCO_hfs_fits_PosInt} we show the temperature dependence of the peak positions resulting from model A  (see Fig. \ref{LayoutSFCOcompareFits}) with broadened hfs lines. The softening of the peak positions with temperature can be fitted by a power law  which serves merely as a guide to the eye. The area of the hfs peaks remains constant up to nearly $T_{sf}$.

\begin{figure}
\resizebox{0.45\textwidth}{!}{\includegraphics{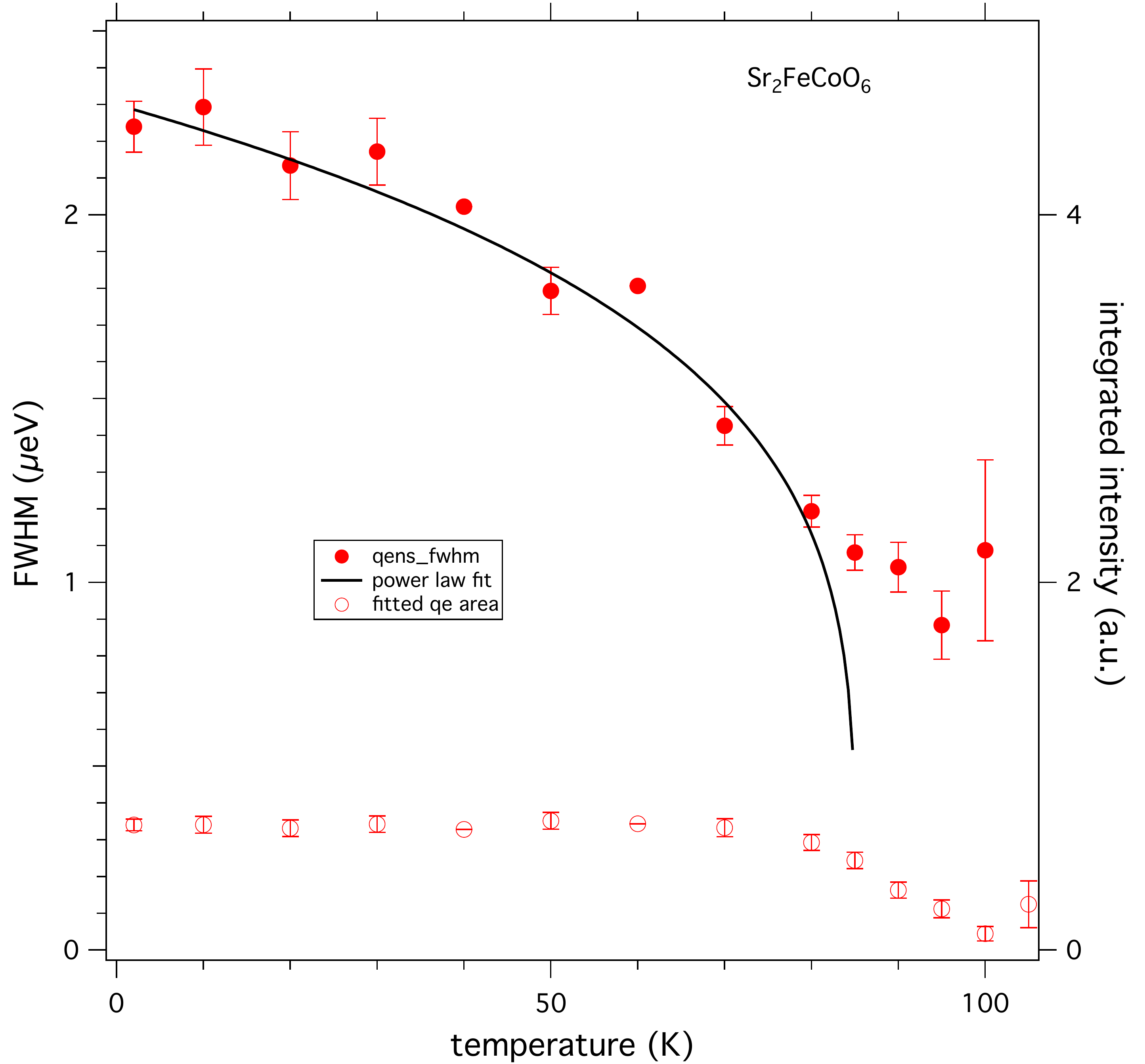}}
\caption {Results from fitting the spectra of Sr$_2$FeCoO$_6$ with model B. Note that the additional elastic scattering is dominating and has e.g. at the lowest temperature a more than a factor of 5 larger area.}
\label{SFCO_qens_fits_WidthInt}
\end{figure}

The temperature dependence of the fitparameters for fit model B (see Fig. \ref{fit_hfsLorE_SLFCO5b01_2K}), i.e. with a quasielastic component, result in a linewidth which decreases with temperature as shown in Fig.\ref{SFCO_qens_fits_WidthInt}. Again the observed narrowing can be fitted by a power law which serves merely as a guide to the eye. Like with model A the area of the hfs peaks remains constant up to nearly $T_{sf}$.
 
\begin{figure}
\resizebox{0.4\textwidth}{!}{\includegraphics{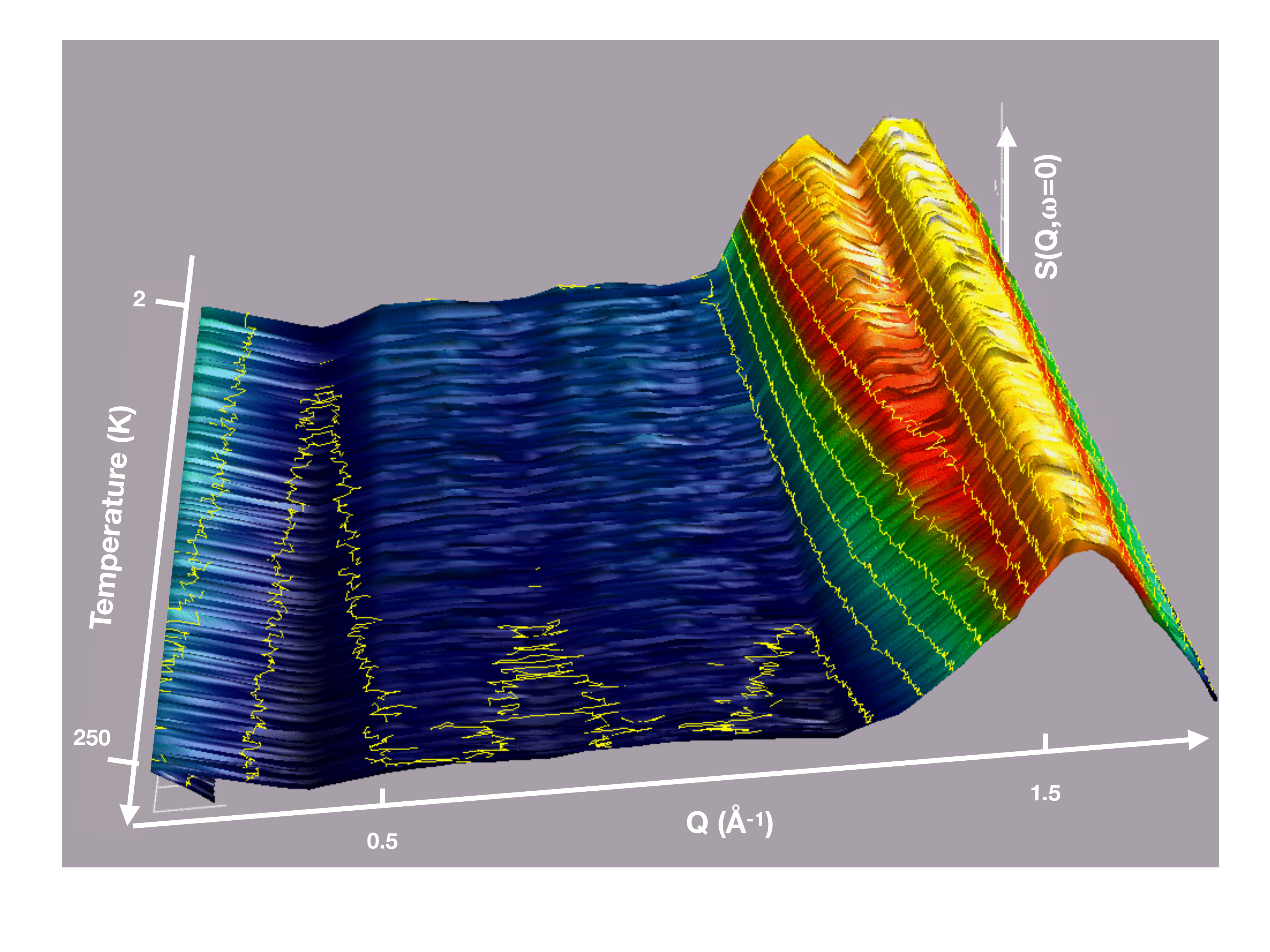}}
\caption {Elastic intensity of La$_2$FeCoO$_6$ as a function of temperature (increasing from back to front) and Q as measured on IN16B. The maximum intensity at the r.h.s. corresponds to a Bragg peak at Q $\approx$ 1.61 {\AA}$^{-1}$ and in cooling below T = 225 K its shoulder at the low Q side develops into a further Bragg peak near Q $\approx$ 1.4 {\AA}$^{-1}$.}
 \label{LFCO_efws_3D}
\end{figure}

\subsection{La$_2$FeCoO$_6$}

For the magnetically ordered system La$_2$FeCoO$_6$ we have shown in Fig.\ref{LFCOefws} the existence of an additional Bragg peak at low temperatures. The elastic fixed window scan for {La$_2$FeCoO$_6$} in Fig.\ref{LFCO_efws_3D} shows how this additional Bragg peak develops continuously with decreasing temperature near Q $\approx$ 1.4 {\AA}$^{-1}$ within of the low Q-shoulder of the Q=1.61 {\AA}$^{-1}$ Bragg peak.


\chapter{References}

\begin{thebibliography}{99}
\bibitem{heidemann_70} A. Heidemann, Z. Physik {\bf238}, 208 (1970).
\bibitem{heidemann_CoP} A. Heidemann, Z. Phys. B {\bf20}, 385 (1975).
\bibitem{heidemann72} A. Heidemann abd B. Alefeld, V. IAEA Symp. Inelastic Neutron Scattering, Grenoble 1972, IAEA-SM-155/G-4.
\bibitem{chatterji00}T. Chatterji and B. Frick, Physica B {\bf 276 - 278}, 252 (2000).
\bibitem{chatterji12}T. Chatterji, J. Combet, B. Frick, and A. Szyula, J. Magn. Magn. Mater. {\bf 324}, 1030 (2012).
\bibitem{chatterji10}T. Chatterji, J. Wuttke and A.P. Sazonov, J. Magn. Magn. Mater. {\bf 322}, 3148 (2010).
\bibitem{chatterji13}T. Chatterji and N. Jalarvo, J. Phys.: Condens. Matter {\bf 25}, 156002 (2013).
\bibitem{chatterji13a}T. Chatterji, N. Jalarvo, C.M.N. Kumar, Y. Xiao and T. Br\"uckel,  J. Phys.: Condens. Matter {\bf 25}, 286003 (2013).

\bibitem{tunneling} J. Colmenero, R. Mukhopadhyay, A. Alegria, and B. Frick, Phys. Rev. Lett. {\bf 80}, 2350 (1998).

\bibitem{garcia01}M. Garc\'ia-Herm\'andez, J. Mart\'inez, M. Mart\'inez-Lope, M. Casais and J.A. Alonso, Phys. Rev. Lett. {\bf 86}, 2443 (2001)
\bibitem{carlo11}J. P. Carlo, J. P. Clancy, T. Aharen, Z. Yamani, J. P. C. Ruff, J. J. Wagman, G. J. Van Gastel, H. M. L. Noad, G. E. Granroth, J. E. Greedan, H. A. Dabkowska, and B. D. Gaulin, Phys. Rev. B {\bf 84}, 100404 (2011).
\bibitem{du10}Y. Du, Z.X. Cheng, S.X. Dou, X.L. Wang, H.Y. Zhao and H. Kimura, Appl. Phys. Lett. {\bf 97}, 122502 (2010).
\bibitem{rogado05}N.S. Rogado, J. Li, A.W. Sleight, M.A. Subramanian, Adv. Mater. {\bf 17}, 2225-7 (2005)


\bibitem{pradheesh12}Pradheesh R, Nair H S, Kumar C M N, Lamsal J, Nirmala R, Santhosh P N, Yelon W B, Malik S K, Sankaranarayanan Vand Sethupathi K  J. Appl. Phys. {\bf 111} 053905 (2012).
 \bibitem{pradheesh12epj}R. Pradheesh, H.S. Nair, V. Sankaranarayanan, K. Sethupathi, Eur. Phys. J. B {\bf85},8 (2012)
\bibitem{pradheesh17} R Pradheesh, H. S.  Nair, G R Haripriya, A. Senyshyn,T. Chatterji, V Sankaranarayanan and K Sethupathi, J. Phys.: Condens. Matter {\bf 29}, 095801(2017).
\bibitem{pradheesh18} Pradheesh R., C.M.N. Kumar, Haripriya G.R., L.M. Martinez, C.L. Caiz, S.S. Rao, T. Chatterji, V. Sankaranarayanan, K. Sethupathi and H.S. Nayer (To be published).

\bibitem{wuttke}J. Wuttke et al., Review of Scientific Instruments, {\bf83}, 075109 (2012).
\bibitem{in16b} http://www.ill.eu/instruments-support/instruments-groups/instruments/in16b/characteristics/
 as well as B. Frick, M. Appel, T. Seydel, L. van Eijck and D. Bazzoli, in preparation.
 \bibitem{in16bexperimentDOI} ILL experiment carried out under DOI: 10.5291/ILL-DATA.4-04-482.

\bibitem{appel}M. Appel and B. Frick, Rev. Sci. Instrum. {\bf88}, 036105 (2017).
 \bibitem{lamp} LAMP standard ILL data evaluation software

\bibitem{chatterji} T. Chatterji et. al. (unpublished results)
\bibitem{mondelli} C. Mondelli et al., Physica B {\bf266}, 104 (1999)
\bibitem{frick_ifws} B. Frick, J. Combet and L. van Eijck, Nucl. Instrum. Meth. A {\bf669}, 7 (2012).
\bibitem{fuh} H.-R. Fuh, K.-C. Weng, C.-R. Chang, and Y.-K. Wang, J. Appl. Phys. {\bf117}, 17B902 (2015).
\bibitem{labrim} H. Labrim, A. Jabar, A. Belhaj, S. Ziti, L. Bahmad, L. La‰nab, and A. Benyoussef, J Alloy Compd {\bf641}, 37 (2015).
\bibitem{chatterji09}T. Chatterji and G.J. Schneider, J. Phys.: Condens. Matter {\bf 21}, 436008 (2009).
\end{thebibliography}
\end{document}